\documentclass[aps,twocolumn,showpacs,pra,floatfix,superscriptaddress]{revtex4-1}
\usepackage{multirow,amssymb,amsbsy,amsmath}
\usepackage{graphicx}
\usepackage{verbatim}
\usepackage[colorlinks=true,linkcolor=blue,citecolor=blue]{hyperref}
\usepackage{subfigure}
\usepackage{braket}

\begin{document}

\title{Generic features of the dynamics of complex open quantum systems: \\
Statistical approach based on averages over the unitary group}





\author{Manuel Gessner}

\email{manuel.gessner@physik.uni-freiburg.de}

\affiliation{Physikalisches Institut, Universit\"at Freiburg,
Hermann-Herder-Strasse 3, D-79104 Freiburg, Germany}

\affiliation{Department of Physics, University of California, Berkeley, California 94720, USA}

\author{Heinz-Peter Breuer}

\email{breuer@physik.uni-freiburg.de}

\affiliation{Physikalisches Institut, Universit\"at Freiburg,
Hermann-Herder-Strasse 3, D-79104 Freiburg, Germany}

\date{\today}

\begin{abstract}
We obtain exact analytic expressions for a class of functions expressed as 
integrals over the Haar measure of the unitary group in $d$ dimensions. Based on 
these general mathematical results, we investigate generic dynamical properties of 
complex open quantum systems, employing arguments from ensemble theory. 
We further generalize these results to arbitrary eigenvalue distributions, allowing 
a detailed comparison of typical regular and chaotic systems with the help of 
concepts from random matrix theory. To illustrate the physical relevance and the 
general applicability of our results we present a series of examples related to the 
fields of open quantum systems and nonequilibrium quantum thermodynamics. 
These include the effect of initial correlations, the average quantum dynamical 
maps, the generic dynamics of system-environment pure state entanglement and, 
finally, the equilibration of generic open and closed quantum systems.
\end{abstract}


\maketitle

\section{Introduction}
The theory of open quantum systems allows for the description of quantum 
systems interacting with their environment \cite{BREUERBOOK}. Recently, the 
focus of research has been drawn towards more and more complex models and 
even biological systems have been approached by the methods of open systems 
\cite{ISHIZAKI,PLENIO,SHABANI}. Increasing size and complexity lead to more 
interesting and realistic models but often cause the treatment of the problem to 
become very cumbersome, if not impossible. In this paper, instead of investigating 
a special system, we are looking for generic features which emerge in a variety of 
complex systems. We study these features employing a statistical approach based 
on averages over a multitude of realizations as it is usual practice in ensemble 
theory for statistical physics \cite{PATHRIA}. Similar approaches have already 
been applied to complex quantum systems 
\cite{WINTER,ZNIDARIC2012,HORODECKI,TIERSCH,GARCIAMATA,HASTINGS,GB}. We analyze dynamical features of open systems by averaging over the unitary 
time evolution operator, which determines the dynamics of system and 
environment. The thereby obtained results reflect the behavior of a generic 
complex open quantum system.

Investigating the generic dynamical properties of complex open quantum 
systems through such a statistical approach one encounters statistical
averages of certain operators which can be expressed as integrals over the 
uniform Haar measure on the unitary group. We will employ advanced methods 
from group and representation theory to carry out these integrals.
All integrals of interest in this paper can be written as functions of the form
\begin{align} \label{E-N}
 \mathbb{E}^{(n)}(X_1,\dots,X_{n-1})=\int d\mu(U) U^{\dagger}X_1U\dots U^{\dagger}X_{n-1}U,
\end{align}
where $X_1,\dots,X_{n-1}$ are arbitrary operators on a $d$-dimensional Hilbert 
space $\mathcal{H}$ and $d\mu(U)$ denotes the Haar measure on the 
unitary group $\mathcal{U}(d)$. The expression on the right-hand side
of Eq.~(\ref{E-N}) involves an integration over $n$ unitary matrices and 
in the following the function $\mathbb{E}^{(n)}(X_1,\dots,X_{n-1})$ will be called 
the \textit{$n$th moment function of the unitary group}. The results of Collins and 
\'Sniady \cite{COLLINS} allow, in principle, for an exact determination 
of these moments up to arbitrary order. In this paper, moments up to the order of 
$n=8$ will appear. For the sake of readability, the explicit derivation of these 
expressions is shifted to the appendix.

Equipped with these mathematical tools we are able to analytically tackle a variety 
of physically motivated expressions. In Sec.~\ref{ch.generalresults}, 
we present three general results which will form the foundation for most of the 
following analysis in this paper. These results involve the expectation value of the 
norm of an arbitrary operator, which has been subjected to a unitary 
time evolution before the environment is traced out. We obtain a uniform average, 
as well as the corresponding variance. Additionally, a decomposition of the 
Hamiltonian into eigenvalues and eigenvectors allows to distinguish between the 
generic effects of typical systems with well-known energy-level distributions, which 
will be put into context with random matrix theory and quantum chaos 
\cite{HAAKE,MEHTA,GARCIAMATA,GB}.

In subsequent sections, the versatility of the general results is demonstrated by a 
study of various applications which are of current interest in the fields of open 
quantum systems \cite{BREUERBOOK}, quantum information theory 
\cite{NIELSEN}, quantum chaos \cite{HAAKE,MEHTA} and quantum 
thermodynamics \cite{GEMMER}. In Sec.~\ref{ch.correlationsbipartite} we 
investigate the generic signature of system-environment correlations in the open-system dynamics. In Sec.~\ref{ch.genericevolution} we identify the average 
reduced dynamics of an open quantum system as the depolarizing quantum 
channel. Section \ref{ch.SEentanglement} deals with the average system-environment pure state entanglement dynamics. Finally, in 
Sec.~\ref{ch.equilibration}, we apply our approach to the thermalization process 
of generic complex quantum systems, examining both isolated and open systems.

\section{Average Hilbert-Schmidt norm evolution in an open quantum system}
\label{ch.generalresults}

We consider an open quantum system, composed of the Hilbert spaces of system $\mathcal{H}_S$ and environment $\mathcal{H}_E$ with dimensions $d_S$ and $d_E$, respectively. The composition of system and environment is assumed to be closed, such that we obtain unitary dynamics for the total system. Furthermore, we denote by $M$ an arbitrary, fixed self-adjoint operator with spectral decomposition $M=\sum_{i=1}^dm_i\ket{i}\!\bra{i}$, acting on the $d$-dimensional Hilbert space $\mathcal{H}=\mathcal{H}_S\otimes\mathcal{H}_E$ of the closed system. In the following we are interested in the open system part of $M$ after being exposed to the unitary time evolution of the total system. More precisely, we introduce the following operator:
\begin{align}
 \Delta_t=\text{Tr}_E\{U_tMU_t^{\dagger}\},
\end{align}
where $U_t$ denotes the unitary time evolution operator, propagating the total system states from time $0$ to $t$, and $\text{Tr}_E$ the partial trace over the environment. 

Our goal in this section is to obtain an expectation value of the squared Hilbert-Schmidt norm of $\Delta_t$, averaging over all unitary time evolution operators $U_t$. We choose the squared Hilbert-Schmidt norm $\|\Delta\|^2=\text{Tr}\{\Delta^{\dagger}\Delta\}$ since it enables us to analytically obtain exact expressions for the expectation values in terms of the functions 
$\mathbb{E}^{(n)}$ defined in Eq.~(\ref{E-N}).

\subsection{The uniform average} \label{uniform-average}
In our first approach we replace the time evolution operator $U_t$ by a random 
matrix $U$ and average over all possible realizations of unitary matrices 
employing the Haar measure $d\mu(U)$. This is an invariant measure, 
giving uniform statistical weights to all unitary matrices. We use the following 
notation for operator-valued averages over the Haar measure:
\begin{align}
\left\langle F(U)\right\rangle=\int d\mu(U)F(U).
\end{align}
This approach leads to time-independent results as the time argument is lost in the averaging process. Later we generalize this approach in a way which allows to maintain the explicit time dependence of the average value.

\subsubsection{The expectation value}
We now derive the unitary average of the function $||\Delta||^2=||\text{Tr}_E\{UMU^{\dagger}\}||^2$. Let $\left\{\ket{\varphi_i}\right\}_{i=1}^{d_S}$ and $\left\{\ket{\chi_j}\right\}_{j=1}^{d_E}$ be fixed orthonormal bases of $\mathcal{H}_S$ and $\mathcal{H}_E$, respectively. The elements of the matrix $\Delta$ can be expressed as
\begin{align}
\Delta_{kl}&=\sum_{i,j}m_i\bra{\varphi_k\chi_j}U\ket{i}\!\bra{i}U^{\dagger}\ket{\varphi_l\chi_j}\notag\\
&=\sum_im_i\bra{i}U^{\dagger}A_{kl}U\ket{i},
\label{eq.deltamatrixelements}
\end{align}
where we have defined the operators $A_{kl}=\ket{\varphi_l}\!\bra{\varphi_k}\otimes I$ with the property $A_{kl}^{\dagger}=A_{lk}$, $I$ denoting the identity on $\mathcal{H}_E$. The squared Hilbert-Schmidt norm of the Hermitian matrix $\Delta$ then reads
\begin{align}
\left\|\Delta\right\|^2&=\text{Tr}\Delta^2=\sum_{k,l}\Delta_{kl}\Delta_{lk}\notag\\
&=\sum_{k,l}\sum_{i,j}m_im_j\bra{i}U^{\dagger}A_{kl}U\ket{i}\!\bra{j}U^{\dagger}A_{kl}^{\dagger}U\ket{j}.
\end{align}
The average value of this expression can be written in terms of the general fourth moment function $\mathbb{E}^{(4)}$ of the unitary group, given in the appendix in Eq.~(\ref{eq.fourthmoment}), with  $X_1=X_3^{\dagger}=A_{kl}$ and $X_2=\ket{i}\!\bra{j}$:
\begin{align}
 \left\langle\|\Delta\|^2\right\rangle=\sum_{k,l}\sum_{i,j}m_im_j\bra{i}\mathbb{E}^{(4)}(A_{kl},\ket{i}\!\bra{j},A_{kl}^{\dagger})\ket{j}.\label{eq.intermsofmap}
\end{align}
Using $d=d_Sd_E$ and the relations $\text{Tr}\{A_{kl}^{\dagger}A_{kl}\}=d_E$ and $\text{Tr} A_{kl}=\text{Tr} A_{kl}^{\dagger}=\delta_{kl}d_E$, we obtain the general result for the uniform average \cite{GB}:
\begin{align}
\left\langle\left\|\text{Tr}_{E}\left\{UMU^{\dagger}\right\}\right\|^2\right\rangle=C_1\left\|M\right\|^2+C_2\left(\text{Tr} M\right)^2\label{eq.HSnormaverage},
\end{align}
with
\begin{align}
 C_1=\frac{d_S^2d_E-d_E}{d_S^2d_E^2-1} \;\;\;
 \text{and} \;\;\; C_2=\frac{d_Sd_E^2-d_S}{d_S^2d_E^2-1}.
\end{align}
Note that in the limit of high environmental Hilbert space dimensions, the coefficient $C_1$ vanishes while $C_2$ approaches the asymptotic value of $1/d_S$.

\subsubsection{The variance}
\begin{figure}
\includegraphics[width=.48\textwidth]{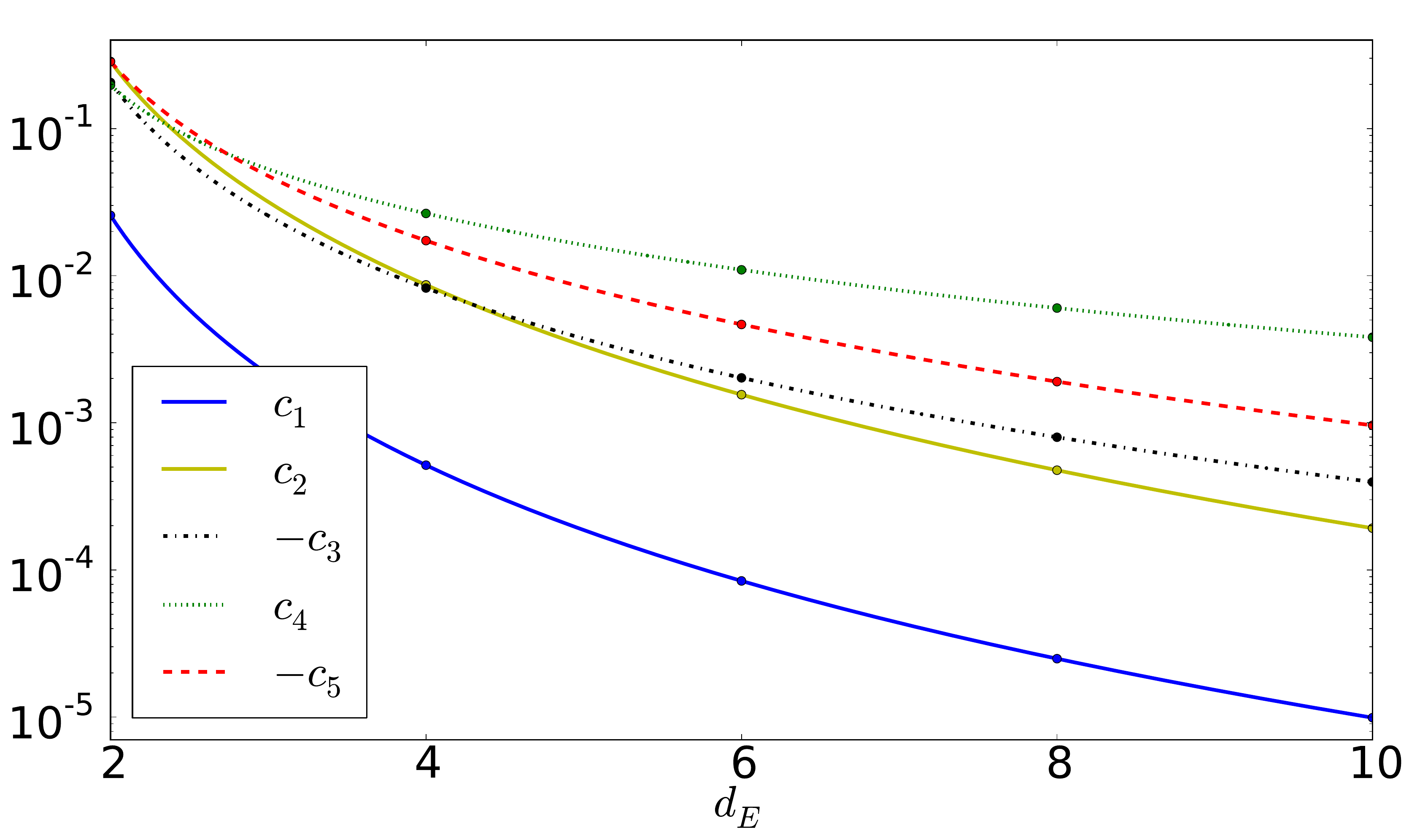} 
\caption{(Color online) Absolute values of the coefficients (\ref{eq.coeffs}) for $d_S=2$ as a function of $d_E$. All coefficients asymptotically tend towards zero in the limit of high environmental dimensions.}
\label{fig.coeffsvariance}
\end{figure}

Next we derive the corresponding variance of the average value (\ref{eq.HSnormaverage}), which is defined as
\begin{align}
 \text{Var}(\|\Delta\|^2)=\left\langle\|\Delta\|^4\right\rangle-\left\langle\|\Delta\|^2\right\rangle^2.\label{eq.variance}
\end{align}
Since the average value is already known, the task is to determine the quantity $\langle\|\Delta\|^4\rangle$. In terms of the matrix elements given in Eq.~(\ref{eq.deltamatrixelements}) this yields
\begin{align}
 \|\Delta\|^4=\:&\sum_{i,j,k,l}\Delta_{ij}\Delta_{ji}\Delta_{kl}\Delta_{lk}\notag\\
=\:&\sum\limits_{\substack{i,j,k,l\\\alpha,\beta,\gamma,\delta}}m_{\alpha}m_{\beta}m_{\gamma}m_{\delta}\bra{\alpha}U^{\dagger}A_{ij}U\ket{\alpha}\notag\\&\times\braket{\beta|U^{\dagger}A_{ji}U|\beta}\!\braket{\gamma|U^{\dagger}A_{kl}U|\gamma}\!\bra{\delta}U^{\dagger}A_{lk}U\ket{\delta}.
\end{align}
Here, eight unitaries are involved in the integration. Hence, we can obtain the result with the aid of the general eighth moment function 
$\mathbb{E}^{(8)}$, given in the appendix in Eq.~(\ref{eq.eighthmoment}), making the following choice of matrices:
\begin{align}
 X_1&=X_3^{\dagger}=A_{ij},\: X_5=X_7^{\dagger}=A_{kl},\notag\\
 X_2&=\ket{\alpha}\!\bra{\beta},\:X_4=\ket{\beta}\!\bra{\gamma},\: X_6=\ket{\gamma}\!\bra{\delta}.
\end{align}
The variance is given according to Eq.~(\ref{eq.variance}),
\begin{align}
 \text{Var}(\|\Delta\|^2)=\:&c_1(\text{Tr} M)^4+c_2(\text{Tr} M)(\text{Tr} M^3)\notag\\&+c_3(\text{Tr} M)^2(\text{Tr} M^2)+c_4(\text{Tr} M^2)^2\notag\\&+c_5(\text{Tr} M^4),\label{eq.generalvariance}
\end{align}
with the coefficients
\begin{align}\label{eq.coeffs}
 c_1&=2b(11+d_S^2d_E^2),\notag\\
 c_2&=40a,\notag\\
 c_3&=-4bd_Sd_E(11+d_S^2d_E^2),\notag\\
 c_4&=2b(15-4d_S^2d_E^2+d_S^4d_E^4),\notag\\
 c_5&=-10ad_Sd_E,
\end{align}
where 
\begin{equation*}
 a=(d_E^2-1)(d_S^2-1)(d_S^2d_E^2(d_S^2d_E^2-7)^2-36)^{-1}, 
\end{equation*}
and 
\begin{equation*}
 b=(d_S^2-1)(d_E^2-1)(d_S^2d_E^2-1)^{-2}(36-13d_S^2d_E^2+d_S^4d_E^4)^{-1}.
\end{equation*}
For increasing environmental dimension $d_E$, all coefficients $c_1,\dots,c_5$ 
tend towards zero, indicating a general universality of the average 
value (\ref{eq.HSnormaverage}) for high Hilbert space dimensions, see 
Fig.~\ref{fig.coeffsvariance}. This result is in agreement with the 
phenomenon of the concentration of measure, which is based on 
L\'evy's lemma \cite{LEVYSLEMMAa,LEVYSLEMMAb}. The decreasing variance 
for this class of average values also endorses the statement that high-dimensional 
generic complex systems are well described by the average value.

\subsection{The average for general eigenvalue distributions}
\subsubsection{The general result}
As a generalization of the result of the previous section, we decompose the Hamiltonian into eigenvectors and eigenvalues as $H=WDW^{\dagger}$, where $W$ denotes a unitary matrix containing the eigenvectors of $H$ and $D=\text{diag}(E_1,E_2,\dots)$ represents a diagonal matrix with the eigenvalues of $H$ as diagonal elements \cite{GARCIAMATA}. The unitary time evolution operator can be written as $U_t=We^{-iDt}W^{\dagger}$. Assuming the eigenvectors to be random and independent of the eigenvalue distribution, we integrate over $W$ with the uniform Haar measure and thereby obtain the generalized version of Eq.~(\ref{eq.HSnormaverage}):
\begin{align}
 \left\langle\|\widetilde{\Delta}\|^2\right\rangle=\int d\mu(W)\left\|\text{Tr}_{E}\left\{We^{-i Dt}W^{\dagger}MWe^{i Dt}W^{\dagger}\right\}\right\|^2,\label{eq.uaveragegeneraleigenvalue}
\end{align}
where we have defined the system matrix $\widetilde{\Delta}=\text{Tr}_{E}\left\{We^{-i Dt}W^{\dagger}MWe^{i Dt}W^{\dagger}\right\}$. 
The time-dependence has been retained within this approach in contrast to the ansatz of Sec.~\ref{uniform-average}. The following procedure is 
however very similar to the previous derivation.
In terms of the matrix elements of $\widetilde{\Delta}$,
\begin{align}
 \widetilde{\Delta}_{kl}=\sum_{\mu}m_{\mu}\braket{\mu|We^{i Dt}W^{\dagger}A_{kl}We^{-i Dt}W^{\dagger}|\mu},
\end{align}
the average value~(\ref{eq.uaveragegeneraleigenvalue}) takes the form
\begin{align}
 &\left\langle\|\widetilde{\Delta}\|^2\right\rangle=\sum_{\mu,\nu,k,l}m_{\mu}m_{\nu}\notag\\&\times\bra{\mu}\mathbb{E}^{(8)}(e^{i Dt},A_{kl},e^{-i Dt},\ket{\mu}\!\bra{\nu},e^{i Dt},A_{lk},e^{-i Dt})\ket{\nu}.\label{eq.complexsystemsum}
\end{align}
We have made use of the general result of the eighth moment function $\mathbb{E}^{(8)}$, see Eq.~(\ref{eq.eighthmoment}), with the special choice of the matrices
\begin{align}
 X_1^{\dagger}&=X_3=X_5^{\dagger}=X_7=e^{-i Dt},\notag\\
 X_2&=X_6^{\dagger}=A_{kl},X_4=\ket{\mu}\!\bra{\nu}.
\end{align}
Finally, we obtain
\begin{align}
 \left\langle\|\widetilde{\Delta}\|^2\right\rangle=\:&\widetilde{c}_{1}(t)\left\|M\right\|^2+\widetilde{c}_{2}(t)\left(\text{Tr} M\right)^2\notag\\&+\widetilde{c}_{3}(t)\left\|\text{Tr}_{E}M\right\|^2+\widetilde{c}_{4}(t)\left\|\text{Tr}_{S}M\right\|^2,\label{eq.generalresultcomplexsystem}
\end{align}
where the coefficients $\widetilde{c}_1(t),\dots,\widetilde{c}_4(t)$ are functions of the dimensions of system and environment and the distribution of the eigenvalues $E_j$ of the Hamiltonian. Introducing the Fourier transform of the level density
\begin{align}
 f(t)=\frac{1}{d}\sum_{j=1}^de^{-i E_jt},
\label{eq.ftld}
\end{align}
these coefficients can be expressed as:
\begin{align}
 \widetilde{c}_1(t)=\:&\big[(d^2-3 d_E^2)b(t)-2d^2(d_E^2-3)\Re\{f(t)^2f^*(2t)\}\notag\\&+(d^2-9)(d^2-d_E^2)\big]/(\widetilde{a}d_E),\notag\\
 \widetilde{c}_2(t)=\:&\big[d(d_E^2-3)b(t)-2d(d^2-3 d_E^2)\Re\{f(t)^2f^*(2t)\}\notag\\&+d(d^2-9)(d_E^2-1)\big]/(\widetilde{a}d_E),\notag\\
 \widetilde{c}_3(t)=\:&-\big[(d^2-3)b(t)+4d^2\Re\{f(t)^2f^*(2t)\}\big]/\widetilde{a},\notag\\                                                                                                                                                                                                                                                     
 \widetilde{c}_4(t)=\:&\big[2db(t)+2d(d^2-3)\Re\{f(t)^2f^*(2t)\}\big]/\widetilde{a},\label{eq.coefficientcomplexsystem}
\end{align}
where
\begin{eqnarray*}
 \widetilde{a} &=& d^4-10 d^2+9, \\ 
 b(t) &=& 4|f(t)|^2-|f(2t)|^2-d^2|f(t)|^4,
\end{eqnarray*}
and $\Re$ denotes the real part of a complex number. We observe that the
coefficients $\widetilde{c}_1(t),\dots,\widetilde{c}_4(t)$ depend only on the 
dimensions of system and environment and on the eigenvalue distribution of the 
Hamiltonian via functions of $f(t)$.

\subsubsection{Averages for regular and chaotic systems}
From the theory of quantum chaos and random matrix theory it is well known that the energy distribution of complex systems typically represents a signature of regular or chaotic behavior \cite{HAAKE,MEHTA}. In order to distinguish between different ensembles, we average the set of eigenvalues of the total system Hamiltonian according to well-known characteristic statistics. Since the average value (\ref{eq.generalresultcomplexsystem}) depends on the energy spectrum via the function $f(t)$, the averages of $|f(t)|^2$, $\Re\{f(t)^2f^*(2t)\}$, and $|f(t)|^4$ with respect to the eigenvalue distributions in question are required. In Appendix \ref{ch.rmt}, we derive the averages of these functions for regular systems, being represented by Poissonian distributed eigenvalue spacings (Poi), and for chaotic systems with no time reversal symmetry as represented by the Gaussian unitary ensemble (GUE). 

\begin{figure}
 \includegraphics[width=.48\textwidth]{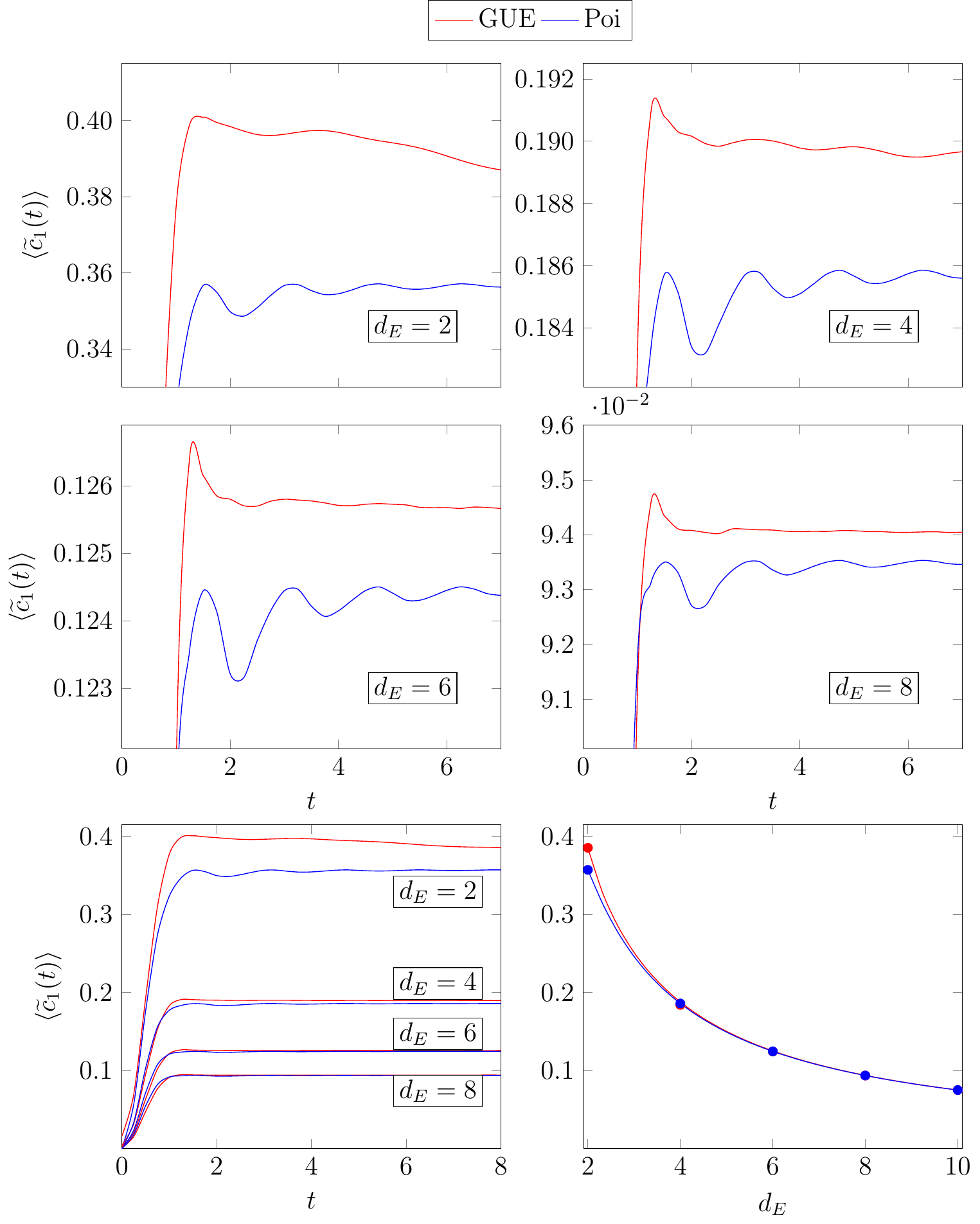}
 \caption{(Color online) Comparison of the average coefficient $\langle \tilde{c}_1(t)\rangle$ for regular (Poi) and chaotic (GUE) systems for $d_S=2$ and different environmental dimensions. The bottom right picture shows the decay of the asymptotic values for $t\rightarrow\infty$ as a function of $d_E$.}
 \label{fig.c1oft}
\end{figure}

As an example we show in Fig.~\ref{fig.c1oft} the average time evolution of the coefficient $\tilde{c}_1(t)$ for regular and chaotic systems. Note that here and in the following the time $t$ is taken to be dimensionless, the relevant unit of time being of the order of the quantity $\hbar/\Delta E$, where $\Delta E$ represents the width of the level density $R_1(E)$ (see Appendix \ref{ch.rmt}). The Heisenberg time is given by $t_{H}=\hbar/\overline{\Delta E}=d\hbar/\Delta E$, with the average level spacing $\overline{\Delta E}=\Delta E/d$. Hence, in dimensionless units the Heisenberg time is equal to the Hilbert space dimension $d$.

\subsubsection{Asymptotic values of the coefficients}
\label{ch.complexsystemscoeffs}
The result of the average value (\ref{eq.generalresultcomplexsystem}) depends crucially on the dimensions of system and environment. In this section we analyze this dependence and present a connection to the uniform average, given in Eq.~(\ref{eq.HSnormaverage}).

For the Poissonian level spacing distribution the coefficients $\widetilde{c}_1(t)$ and $\widetilde{c}_4(t)$ disappear in the limit of large $d_E$ with a fixed value of $d_S$ while the other two coefficients tend towards nonzero functions of time:
\begin{align}\label{eq.asymppoi}
 \lim\limits_{d_E\rightarrow\infty}\langle\widetilde{c}_1(t)\rangle_{\text{Poi}}&=0,\notag\\
 \lim\limits_{d_E\rightarrow\infty}\langle\widetilde{c}_2(t)\rangle_{\text{Poi}}&=\frac{128t^4+4\cos(4t)-\cos(8t)-3}{128d_St^4},\notag
 \end{align}\begin{align}
 \lim\limits_{d_E\rightarrow\infty}\langle\widetilde{c}_3(t)\rangle_{\text{Poi}}&=\left(\frac{\cos(t)\sin(t)}{t}\right)^4,\notag\\
 \lim\limits_{d_E\rightarrow\infty}\langle\widetilde{c}_4(t)\rangle_{\text{Poi}}&=0.
\end{align}
If we additionally take the limit for large times $t$, the coefficients $\widetilde{c}_2(t)$ and $\widetilde{c}_3(t)$ yield the following asymptotic values:
\begin{align}
 \lim\limits_{t\rightarrow\infty}\lim\limits_{d_E\rightarrow\infty}\langle\widetilde{c}_2(t)\rangle_{\text{Poi}}&=\frac{1}{d_S},\notag\\
 \lim\limits_{t\rightarrow\infty}\lim\limits_{d_E\rightarrow\infty}\langle\widetilde{c}_3(t)\rangle_{\text{Poi}}&=0.
\end{align}
This coincides with the time-independent result of the uniform average in the same limit, see Eq.~(\ref{eq.HSnormaverage}).

Now we investigate the Gaussian unitary ensemble. For large $d_E$ and fixed $d_S$ we obtain the following results (see Appendix \ref{ch.gue}):
\begin{align}\label{eq.asympgue}
 \lim\limits_{d_E\rightarrow\infty}\langle\widetilde{c}_1(t)\rangle_{\text{GUE}}&=0,\notag\\
 \lim\limits_{d_E\rightarrow\infty}\langle\widetilde{c}_2(t)\rangle_{\text{GUE}}&=\frac{1}{d_S}\left(1-\left[\frac{J_1(2t)}{t}\right]^4\right),\notag\\
 \lim\limits_{d_E\rightarrow\infty}\langle\widetilde{c}_3(t)\rangle_{\text{GUE}}&=\left[\frac{J_1(2t)}{t}\right]^4,\notag\\
 \lim\limits_{d_E\rightarrow\infty}\langle\widetilde{c}_4(t)\rangle_{\text{GUE}}&=0.
\end{align}
After the additional limit for large $t$, again, we find that the values coincide with the results which we obtained for the Poissonian average and the uniform average:
\begin{align}
 \lim\limits_{t\rightarrow\infty}\lim\limits_{d_E\rightarrow\infty}\langle\widetilde{c}_2(t)\rangle_{\text{GUE}}&=\frac{1}{d_S},\notag\\
 \lim\limits_{t\rightarrow\infty}\lim\limits_{d_E\rightarrow\infty}\langle\widetilde{c}_3(t)\rangle_{\text{GUE}}&=0.
\end{align}
While for the Poissonian statistics the asymptotic values for $t\rightarrow\infty$ depend on $d_S$ and $d_E$ in a complicated way, here we find that within the approximation for large values of $d$ (see Appendix \ref{ch.gue}), the Gaussian average is equal to the time-independent result of the uniform average:
\begin{align}
 \lim\limits_{t\rightarrow\infty}\langle\widetilde{c}_1(t)\rangle_{\text{GUE}}&\approx\frac{d_S^2d_E-d_E}{d_S^2d_E^2-1}=C_1,\notag\\
 \lim\limits_{t\rightarrow\infty}\langle\widetilde{c}_2(t)\rangle_{\text{GUE}}&\approx\frac{d_Sd_E^2-d_S}{d_S^2d_E^2-1}=C_2,\notag\\
 \lim\limits_{t\rightarrow\infty}\langle\widetilde{c}_3(t)\rangle_{\text{GUE}}&\approx0,\notag\\
 \lim\limits_{t\rightarrow\infty}\langle\widetilde{c}_4(t)\rangle_{\text{GUE}}&\approx0.
\end{align}

Finally, we remark that although the dimension of the system Hilbert space is usually quite obvious, from a physical point of view, it is not always clear which  is a suitable dimension for the environment. In many cases, the environment consists of a huge Hilbert space of which only a small part actually affects the system dynamics. The effective environmental dimension, to be inserted as $d_E$ in the formula for the averages, is not always given by the mathematical dimension of the environmental Hilbert space, but rather characterizes a certain subspace of the Hilbert space which plays a role in the system's dynamics. For a given physical model the effective environmental dimension may be hard to obtain and further investigations are certainly necessary in order to propose a clean definition and an applicable and efficient way for its theoretical and experimental determination \cite{GB2}.

\section{Correlations in bipartite systems}
\label{ch.correlationsbipartite}
An important special case of the previously derived general results is obtained if $M=\rho-\rho'$, with two states $\rho$ and $\rho'$ of the closed system, is inserted into Eqs.~(\ref{eq.HSnormaverage}) and (\ref{eq.generalvariance}). The uniform average yields \cite{GB}
\begin{align}
\left\langle\left\|\text{Tr}_{E}\left\{U(\rho-\rho')U^{\dagger}\right\}\right\|^2\right\rangle=\frac{d_S^2d_E-d_E}{d_S^2d_E^2-1}\left\|\rho-\rho'\right\|^2,\label{eq.uniformtwostate}
\end{align}
with the variance
\begin{align}
 \text{Var}(\left\|\text{Tr}_{E}\left\{U(\rho-\rho')U^{\dagger}\right\}\right\|^2)=\:&c_4(\text{Tr}\{(\rho-\rho')^2\})^2\notag\\&+c_5\text{Tr}\{(\rho-\rho')^4\}.\label{eq.twostatevariance}
\end{align}

Distance measures are particularly interesting in the context of geometric measures of correlations \cite{MODIREVIEW}. The distance
\begin{align}\label{eq.distance}
\mathcal{D}(\rho)=\|\rho-\rho'\|
\end{align} 
of a given state $\rho$ to an associated uncorrelated state can be interpreted as a measure for the amount of correlations in $\rho$. While in this manuscript we work with the Hilbert-Schmidt distance, other works have focussed on different operator distances and pseudo-distances with appealing properties, like, e.g. the trace-distance, fidelity and the relative entropy \cite{NIELSEN,HHHH,MODIREVIEW}.

A possible measure for the total correlations of a state $\rho$ is the distance of the state $\rho$ to the completely uncorrelated product state of its marginals $\rho'=\rho_S\otimes\rho_E$, where $\rho_S=\text{Tr}_E\rho$ and $\rho_E=\text{Tr}_S\rho$ denote the reduced density operators of $\rho$ for system and environment, respectively \cite{WITNESS}. 

Quantum correlations expressed by the quantum discord \cite{VEDRAL,ZUREK,MODIREVIEW} can be quantified through $\mathcal{D}$ by creating a reference state $\rho'$ with a local dephasing operation applied to the original state $\rho$ \cite{GB,GB2}. This operation describes full dephasing in the eigenbasis of the reduced density matrix and removes all quantum discord of the total state without perturbing the local states. For further details we refer the reader to the literature on this topic \cite{GB,MODIREVIEW,TURKU,GB2}.

From Eq.~(\ref{eq.uniformtwostate}), one can see the remarkable result that the average squared distance of the two reduced states $\rho_S(t)=\text{Tr}_{E}\{U_t\rho U_t^{\dagger}\}$ and $\rho'_S=\text{Tr}_{E}\{U_t\rho'U_t^{\dagger}\}$ is proportional to the squared distance of the initial states of the composite system. The above examples for $\rho$ and $\rho'$ then show that the total correlations or the quantum discord, given by Eq.~(\ref{eq.distance}) have direct impact on the generic evolution in the open system. This can be used in order to detect properties of the initial states locally \cite{GB,GB2}. 

This direct proportionality can also be observed for the average for general eigenvalue distributions whenever the marginal states of $\rho$ and $\rho'$ are equal. More precisely, from Eq.~(\ref{eq.generalresultcomplexsystem}) we find
\begin{align}\label{eq.generaldistance}
&\left\langle\left\|\text{Tr}_{E}\left\{We^{-i Dt}W^{\dagger}(\rho-\rho')We^{i Dt}W^{\dagger}\right\}\right\|^2\right\rangle\notag\\&\;=\widetilde{c}_{1}(t)\left\|\rho-\rho'\right\|^2,
\end{align}
if $\text{Tr}_E\{\rho-\rho'\}=0$ and $\text{Tr}_S\{\rho-\rho'\}=0$. This is obviously the case if $\rho'$ is chosen to be the corresponding product state of $\rho$, but also holds for the locally dephased reference state, as described before \cite{GB,GB2}. The function $\widetilde{c}_1(t)$ is plotted in Fig.~\ref{fig.c1oft} for typical regular and chaotic systems. Equations~(\ref{eq.asymppoi}) and (\ref{eq.asympgue}) show that in both cases, this function approaches zero for large $d_E$. The same holds for the coefficient in the uniform case, displayed in Eq.~(\ref{eq.uniformtwostate}). Hence, we find that the average evolution of an arbitrary pair of quantum states in systems interacting with high-dimensional environments will be increasingly harder to distinguish. We can conclude that in such cases the effect of initial correlations has vanishing influence on the open-system evolution.

\section{Average open-system time evolution: Depolarizing channel}
\label{ch.genericevolution}
In this section we consider the average time evolution of an open quantum system. Making use of the decomposition of the unitary time evolution operator $U_t=W\exp\{-i D t\}W^{\dagger}$, we find for the average reduced system state
\begin{align}
 \left\langle\rho_S(t)\right\rangle=\text{Tr}_{E}\{\underbrace{\left\langle W\exp\{-i D t\}W^{\dagger}\rho(0)W\exp\{i D t\}W^{\dagger}\right\rangle}_{\left\langle\rho(t)\right\rangle}\}
\end{align}
where $\rho(0)$ denotes the initial state of system and environment. The averaged expression contains four unitary matrices and can therefore be obtained with the aid of the general result (\ref{eq.fourthmoment}), with $X_1=X_3^{\dagger}=\exp\{-i D t\}$ and $X_2=\rho(0)$. This yields
\begin{align}
\left\langle\rho(t)\right\rangle&=\left\langle W\exp\{-i D t\}W^{\dagger}\rho(0)W\exp\{i D t\}W^{\dagger}\right\rangle\notag\\&=\frac{d^2-d^2|f(t)|^2}{d^2-1} I/d+\frac{d^2|f(t)|^2-1}{d^2-1}\rho(0)\label{eq.genericevolution}.
\end{align}
In quantum information theory this map $\rho(0)\rightarrow\langle\rho(t)\rangle$ is known as a depolarizing channel \cite{NIELSEN}. For the reduced dynamics we obtain the same map,
\begin{align}
 \left\langle\rho_S(t)\right\rangle=\frac{d^2-d^2|f(t)|^2}{d^2-1} I/d_S+\frac{d^2|f(t)|^2-1}{d^2-1}\rho_S(0),
\end{align}
with the initial reduced state $\rho_S(0)=\text{Tr}_{E}\rho(0)$. Note that in contrast to the usual experience with open quantum systems, this map is always well-defined and completely positive, depending only on the initial reduced state $\rho_S(0)$ regardless of the initial correlations which may be present in $\rho(0)$ \cite{WITNESS}. 

This map has been studied previously in Ref. \cite{GARCIAMATA} where the non-Markovianity of this quantum channel was investigated as a function of the level statistics. A detailed analysis of the statistical implications for the reduced density matrix at the outcome of this channel was carried out in Ref. \cite{ZNIDARIC2012}.

\section{Generic dynamics of system-environment pure state entanglement}
\label{ch.SEentanglement}
As another application of the results of Eqs.~(\ref{eq.HSnormaverage}) and (\ref{eq.generalresultcomplexsystem}), we consider the special case of $M=\rho$, where $\rho$ is an arbitrary fixed state of the total system. The term on the left-hand side of these equations then yields the unitary average value of the purity of the reduced system states. 

The purity $\mathcal{P}(\rho_S)$ is defined as $\mathrm{Tr}(\rho_S^2)$. It attains the maximal value of 1 if and only if the state $\rho_S$ is pure. The minimal value is given by $1/d_S$ and is reached for the maximally mixed state $\rho_S=I/d_S$. The reduced purity is an especially interesting observable in the context of quantum correlations since it reveals the degree of entanglement in pure bipartite states. A maximally mixed state in the subsystem corresponds to a maximally entangled state in the closed system, while a pure reduced state corresponds to a separable total state \cite{SCH35,HHHH}. Many studies on properties of random states have focussed on purity \cite{Gorin1,Giraud1}, entanglement \cite{ZnidaricA,Giraud2,ZnidaricB,Majumdar,Nadal,Kumar} and the connection to decoherence mechanisms \cite{Gorin1}.

By Eq.~(\ref{eq.HSnormaverage}) we obtain the time-independent value for a uniform average, to be discussed in the next section, while the result of Eq.~(\ref{eq.generalresultcomplexsystem}) depends on time and on the eigenvalue distribution and will be discussed in Sec.~\ref{ch.purityrandomensemble}.

\begin{figure}
 \centering
 \includegraphics[width=.48\textwidth]{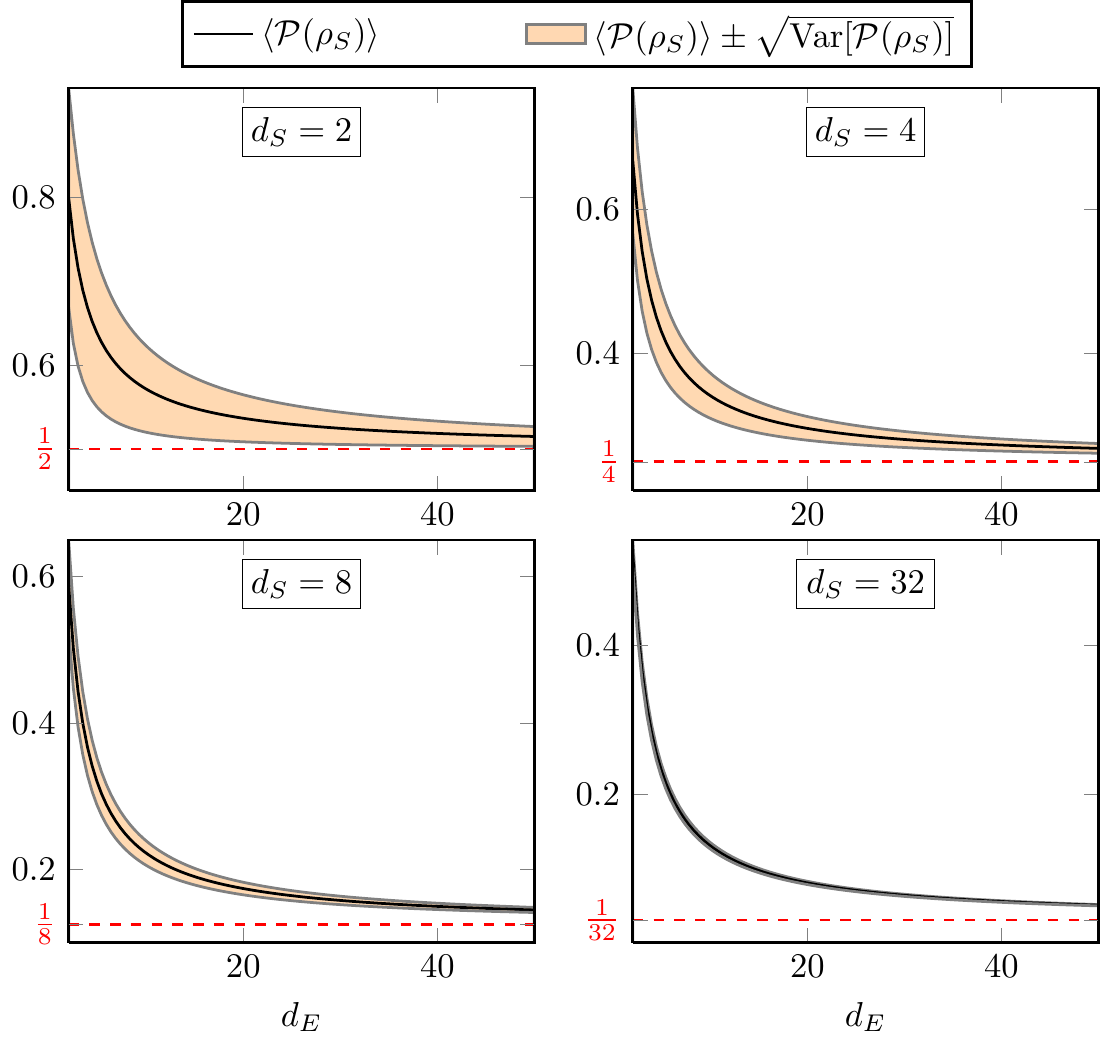}
 \caption{(Color online) Average open-system purity within one standard deviation as a function of $d_E$ for different values of $d_S$. The asymptotic value is given by $1/d_S$. Absolute and relative error decrease for increasing dimensions.}
 \label{fig.purityvariance}
\end{figure}

\subsection{The uniform average}
In the case of a uniform average over the unitary group we get the following connection between the average value and the initial purity:
\begin{align}
\left\langle\mathcal{P}(\rho_S)\right\rangle=\frac{(d_S^2d_E-d_E)\mathcal{P}(\rho)+d_Sd_E^2-d_S}{d_S^2d_E^2-1}\label{eq.genericpurityCUE}.
\end{align}
It is remarkable that the right-hand side of Eq.~(\ref{eq.genericpurityCUE}) only depends on the purity $\mathcal{P}(\rho)$ of the initial state of the closed system and the dimensions of system and environment. In the following we consider a pure state $\rho=\ket{\Psi}\!\bra{\Psi}$ of the closed system with $\mathcal{P}(\rho)=1$. The variance, Eq.~(\ref{eq.generalvariance}), can be simplified to
\begin{align}
 \textrm{Var}[\mathcal{P}(\rho_S)]=\frac{2(d_E^2-1)(d_S^2-1)}{(d_Sd_E+1)^2(d_Sd_E+2)(d_Sd_E+3)}.
\end{align}
For large values of $d_E$, the average purity behaves as
\begin{align}
\left\langle\mathcal{P}(\rho_S)\right\rangle\sim\frac{1}{d_S}+\left(1-\frac{1}{d_S^2}\right)\frac{1}{d_E}.
\end{align}
The asymptotic value $1/d_S$ for $d_E\rightarrow\infty$ corresponds to a maximally mixed state and, hence, to maximal amount of entanglement in the bipartite system. As it was foreshadowed by the general considerations displayed in Fig.~\ref{fig.coeffsvariance} we find that the variance vanishes in the limit of large environmental dimensions, $\textrm{Var}[\mathcal{P}(\rho_S)]\rightarrow0$ for 
$d_E\rightarrow\infty$. The same holds true for the 
relative fluctuations which for large $d_E$ behave as
\begin{align}
 \frac{\sqrt{\textrm{Var}[\mathcal{P}(\rho_S)]}}{\left\langle\mathcal{P}(\rho_S)
 \right\rangle} \sim \sqrt{\frac{2(d_S^2-1)}{d_S^2}} \frac{1}{d_E}.
\end{align}
For high dimensions the pure state entanglement between system and environment after the time evolution with respect to an average unitary operator demonstrates a universal behavior. The asymptotic value depends only on the dimensions of system and environment: For large $d_E$ the total system approaches the maximally entangled state while the fluctuations around the average value decrease to zero. This confirms the conclusion that generic states of high-dimensional systems contain large amounts of entanglement \cite{HAYDEN,ZNIDARIC,PLENIO2,TIERSCH,ZANARDI,VEDRAL2}. Fig.~\ref{fig.purityvariance} shows the dependence of the average purity within one standard deviation on the environmental dimension for four different system sizes.

\begin{figure}
\centering
 \includegraphics[width=.48\textwidth]{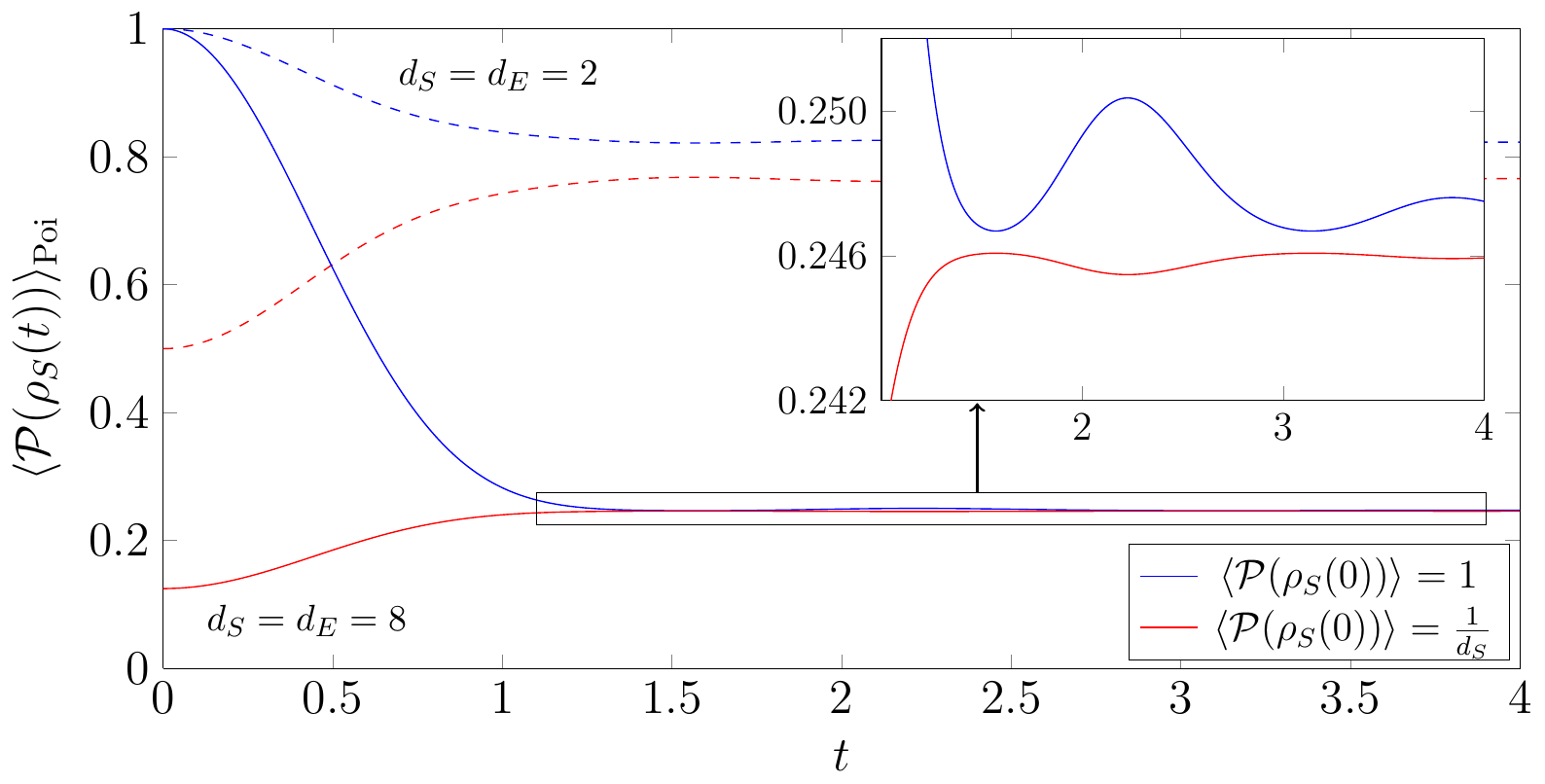}
\caption{(Color online) Generic subsystem purity evolution in a regular system for a pure initial state of the total system. All average values refer to Poissonian statistics. The generic evolution is plotted for two different systems, each for the examples of an initially pure and a maximally mixed state of the reduced system. After a few oscillations the asymptotic value is reached. The dependence of this value on the initial reduced purity vanishes for larger dimensions.}
\label{fig.purityPoiA}
\end{figure}

\subsection{The average for general random matrix ensembles}
\label{ch.purityrandomensemble}
By inserting $M=\rho=\ket{\Psi}\!\bra{\Psi}$ into Eq.~(\ref{eq.generalresultcomplexsystem}), we get a time-dependent average of the reduced state's purity,
\begin{align}
 \langle\mathcal{P}(\rho_S(t))\rangle=\widetilde{c}_1(t)+\widetilde{c}_2(t)+(\widetilde{c}_3(t)+\widetilde{c}_4(t))\langle\mathcal{P}(\rho_S(0))\rangle,\label{eq.genericentanglementdynamics}
\end{align}
which additionally depends on the initial subsystem purity $\langle\mathcal{P}(\rho_S(0))\rangle$. This stems from the terms $\|\text{Tr}_{E}\rho\|^2$ and $\|\text{Tr}_{S}\rho\|^2$, which are equal for a pure state. The average corresponds to an average over the eigenvectors using the Haar measure while the eigenvalue distribution is implicitly included in the constants $\widetilde{c}_1(t),\dots,\widetilde{c}_4(t)$. Since a pure state remains pure under unitary evolution, the above equation characterizes the entanglement between system and environment for any time $t$. Hence, Eq.~(\ref{eq.genericentanglementdynamics}) can be interpreted as the generic system-environment pure state entanglement evolution.

After additionally averaging the coefficients with respect to a certain eigenvalue distribution, symbolically represented by the subscript EVD, the subsystem purity evolution is given by
\begin{align}
 &\langle\mathcal{P}(\rho_S(t))\rangle_{\text{EVD}}=\frac{d_S+d_E}{d_Sd_E+1}\notag\\&\quad+\frac{(d_S+d_E)(\langle g(t)\rangle_{\text{EVD}}-2d_Sd_E\langle\Re\{f(t)^2f^*(2t)\}\rangle_{\text{EVD}})}{(d_Sd_E-1)(d_Sd_E+1)(d_Sd_E+3)}\notag\\
&\quad+\frac{2d_Sd_E\langle\Re\{f(t)^2f^*(2t)\}\rangle_{\text{EVD}}-\langle g(t)\rangle_{\text{EVD}}}{(d_Sd_E-1)(d_Sd_E+3)}\langle\mathcal{P}(\rho_S(0))\rangle,\label{eq.complexsubsystempurity}
\end{align}
with $g(t)=4|f(t)|^2-|f(2t)|^2-d_S^2d_E^2|f(t)|^4$. The average values of these coefficients can now be obtained by inserting the respective average values of the functions of $f(t)$. The derivations of these are given in Appendix \ref{ch.rmt}.

We start with an analysis of regular systems, described by Poissonian level statistics. Inserting the corresponding averages into Eq.~(\ref{eq.complexsubsystempurity}) yields the generic pure state entanglement evolution in a regular complex open system. This function is plotted in Fig.~\ref{fig.purityPoiA} for bipartite systems of equal sizes. We see that the reduced purity reaches an asymptotic value after some transient time $t$. This value depends on the initial subsystem purity but the dependence weakens for high values of $d_E$. The corresponding limit yields the same result as for the uniform average:
\begin{align}
 \lim\limits_{d_E\rightarrow\infty}\lim\limits_{t\rightarrow\infty}\langle\mathcal{P}(\rho_S(t))\rangle_{\text{Poi}}=\frac{1}{d_S}.
\end{align}

\begin{figure}
\begin{center}
\subfigure[Poissonian distribution]{%
\label{fig.purityPoi}
\includegraphics[width=0.245\textwidth]{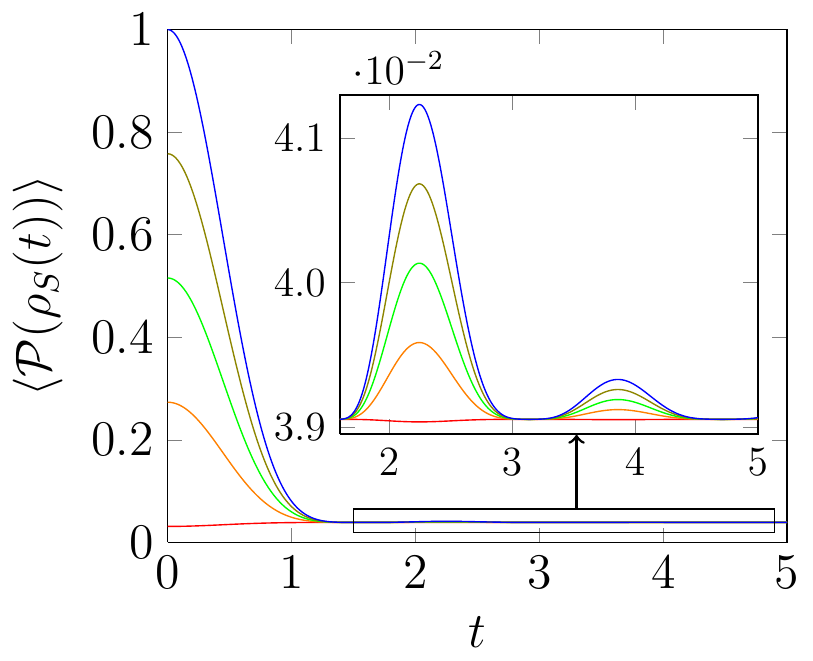}
 }%
\subfigure[Gaussian unitary ensemble]{%
\label{fig.purityGUE}
\includegraphics[width=0.225\textwidth]{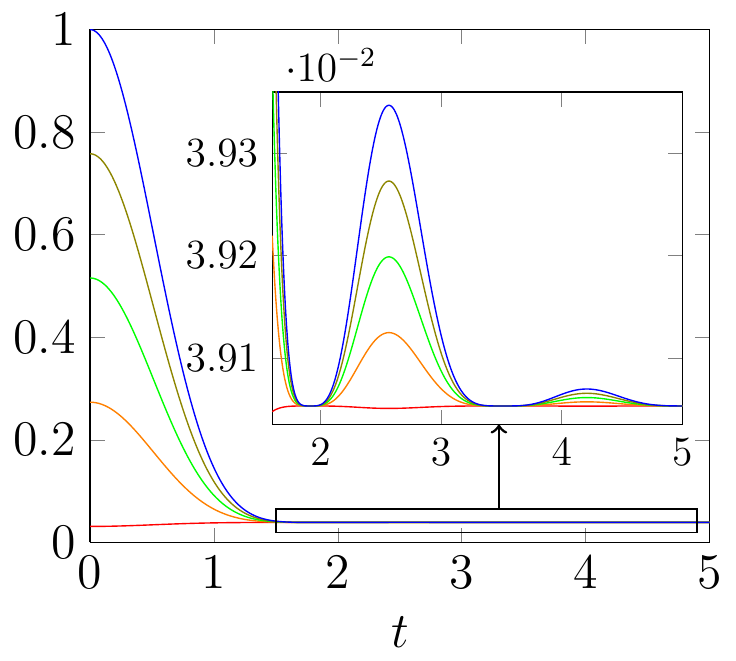}
 }%
\end{center}
\caption{(Color online) Dependence of the subsystem purity evolution on the initial value. The plot shows a system with dimensions $d_S=32$ and $d_E=128$ and five different equidistant values of the initial purity between a pure and a maximally mixed state. \textit{Inlays:} Magnification of the oscillations, which for the GUE are one order of magnitude smaller than for the Poissonian distribution.}
\label{fig.purity32x128}
\end{figure}

Next, we investigate the generic evolution of the same quantity for typical chaotic systems, represented by the Gaussian unitary ensemble. In order to simplify calculations, we employ an approximation for large values of $d$ which is introduced in the Appendix \ref{ch.gue}. Then, the asymptotic value for large times $t$ becomes independent of the initial subsystem purity,
\begin{align}
 \lim\limits_{t\rightarrow\infty}\langle\mathcal{P}(\rho_S(t))\rangle_{\text{GUE}}\approx\frac{d_S+d_E}{d_Sd_E+1}.
\end{align}
An additional limit for large $d_E$ leads to the same result as the Poissonian and uniform averages,
\begin{align}
 \lim\limits_{t\rightarrow\infty}\lim\limits_{d_E\rightarrow\infty}\langle\mathcal{P}(\rho_S(t))\rangle_{\text{GUE}}=\frac{1}{d_S},
\end{align}
The dependence on the initial subsystem purity is elucidated in Fig.~\ref{fig.purity32x128} for a high-dimensional system. Since the Hilbert space dimension is large, all initial states evolve towards the same asymptotic value close to the maximally mixed state for both averages---a dependence on the initial state can only be observed for small times. The overall shape of the two average evolutions is very similar although the oscillations are less pronounced for the GUE.

\begin{figure}[t!]
\centering
 \includegraphics[width=.48\textwidth]{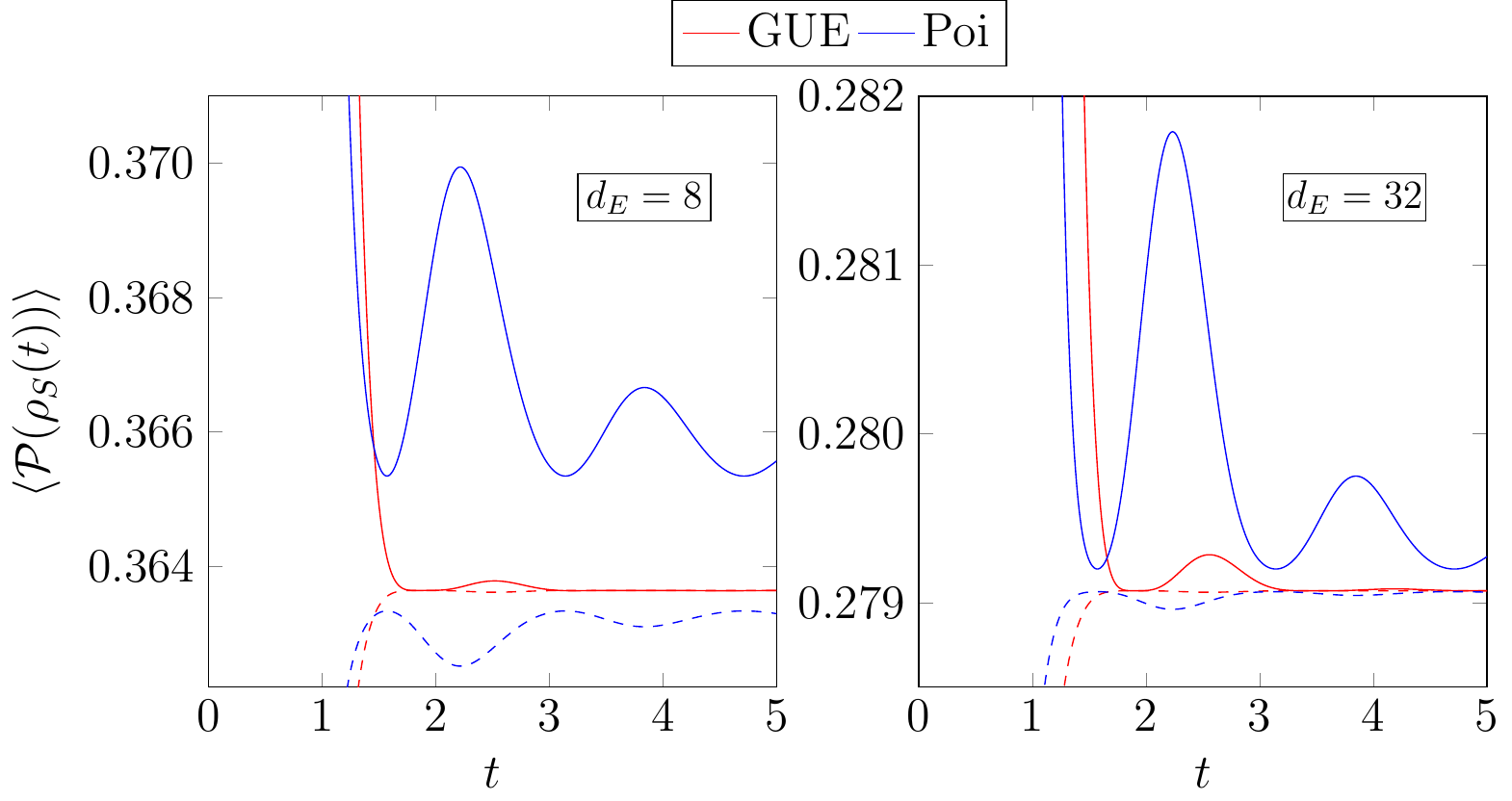}
\caption{(Color online) A close up comparison of the average purity evolution is shown for $d_S=4$ and different environmental dimensions. The continuous lines correspond to the initially pure reduced state and the dashed lines represent the evolution of the initially maximally mixed reduced state.}
\label{fig.purityGUEPoi}
\end{figure}

A comparison of the generic evolution between Poissonian eigenvalue statistics and the GUE is shown in Fig.~\ref{fig.purityGUEPoi}. The universality for large Hilbert space dimensions can be observed graphically: The gap between the asymptotic values for $t\rightarrow\infty$ of two different initial states vanishes as the environmental dimension increases.

\section{Equilibration of generic quantum systems}
\label{ch.equilibration}
As a final application for the general unitary average values, we study the thermalization process of a complex quantum system -- a problem which has attracted a significant amount of interest recently \cite{GEMMER,DEUTSCH,SREDNICKI,CRAMER,LINDEN,RIGOL,HORODECKI}. Before dealing with the evolution of a generic open system we start out with the simpler case of closed-system dynamics. Observation of the distance between the state at time $t$ and the thermal equilibrium state allows to draw conclusions about the thermalization process. In this chapter we will derive explicit equations for the average evolution of this quantity depending on system parameters and the initial conditions. These can be used to infer whether or not a system will thermalize completely, i.e., the distance gets arbitrarily small for large times. In this case also the average time needed to reach the thermal equilibrium state up to a given proximity can be obtained.

\subsection{Closed system thermalization}
The equilibration process can be quantified by the distance of the state $\rho(t)=U_t\rho(0)U_t^{\dagger}$ at time $t$ to the thermal equilibrium state $\rho_G=e^{-\beta H}/Z$, where $Z=\text{Tr}\{e^{-\beta H}\}$ and $\beta=1/kT$ with the Boltzmann constant $k$ and the temperature $T$. As in previous sections, we rewrite the time evolution operator as $U_t=We^{-iDt}W^{\dagger}$. In this decomposition the thermal equilibrium state can be expressed as $\rho_G=W(e^{-\beta D}/Z)W^{\dagger}$. For the Hilbert-Schmidt distance we get
\begin{align}
 &\left\|\rho_G-\rho(t)\right\|^2\notag\\&=\left\|W(e^{-\beta D}/Z)W^{\dagger}-We^{-iDt}W^{\dagger}\rho(0)We^{iDt}W^{\dagger}\right\|^2\notag\\
&=\left\|W(e^{-\beta D}/Z)W^{\dagger}\right\|^2+\left\|We^{-iDt}W^{\dagger}\rho(0)We^{iDt}W^{\dagger}\right\|^2\notag\\&\quad-2\text{Tr}\left\{W(e^{-\beta D}/Z)W^{\dagger}We^{-iDt}W^{\dagger}\rho(0)We^{iDt}W^{\dagger}\right\}.
\end{align}
Integration over $W$ employing the Haar measure yields
\begin{align}
 \left\langle\left\|\rho_G-\rho(t)\right\|^2\right\rangle=\:&\mathcal{P}(\rho_G)+\mathcal{P}(\rho(0))\notag\\&-2\text{Tr}\{e^{-\beta D}/Z\underbrace{\left\langle W^{\dagger}\rho(0)W\right\rangle}_{I/d}\}\notag\\
=\:&\mathcal{P}(\rho_G)+\mathcal{P}(\rho(0))-2/d,
\end{align}
where we have used the invariance property of the trace under cyclic permutations to eliminate some of the unitary operators. Note that the result does not depend on time any more, even though the time dependence has been preserved in the unitary averaging process. It can immediately be confirmed that the value zero is reached for the case where both states $\rho(0)$ and $\rho_G$ are maximally mixed. This corresponds to a thermal state of infinite temperature and minimal purity of $1/d$. All other cases yield nonzero values since the purity is bounded by $1/d\leq\mathcal{P}(\rho)\leq1$.
\begin{figure}
\includegraphics[width=.48\textwidth]{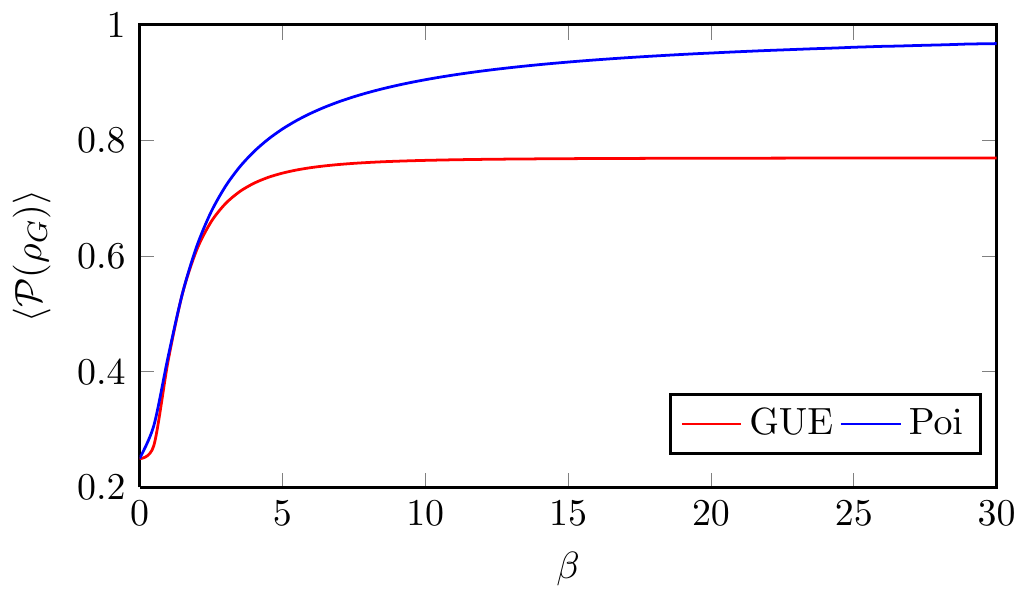}
\caption{(Color online) The dependence of the average thermal purity on the inverse temperature $\beta$ is plotted for Poissonian level spacings and the GUE with $d=4$. Increasing temperature decreases the average purity as the Gibbs state gets closer to the maximal mixture. At low temperatures, regular systems have generically a higher thermal purity than chaotic systems.}
\label{fig.betadependence}
\end{figure}
The purity of the Gibbs state can be expressed in terms of the eigenvalues of $H$:
\begin{align}
 \mathcal{P}(\rho_G)&=\text{Tr}e^{-2\beta D}/(\text{Tr}e^{-\beta D})^2=f(-2i\beta)/f(-i\beta)^2,\label{eq.averagegibbs}
\end{align}
with $f(t)$ as in Eq.~(\ref{eq.ftld}). Due to the negative exponents, the average value of $\mathcal{P}(\rho_G)$ for different eigenvalue distributions cannot be tackled with the same straightforward approach which was used for the previously occuring functions of $f(t)$. For this term we have to explicitly carry out the integration
\begin{align}
 \left\langle\mathcal{P}(\rho_G)\right\rangle=\int dE_1\dots\int dE_dP(E_1,\dots,E_d)\frac{\sum_{i=1}^de^{-2\beta E_i}}{\left(\sum_{i=1}^de^{-\beta E_i}\right)^2},
\end{align}
where $P(E_1,\dots,E_d)$ denotes the probability for the set of energy levels $\{E_1,\dots,E_d\}$, see e.g. \cite{MEHTA}. While for the Poissonian level spacing the energy levels are uniformly distributed, we get for the GUE \cite{MEHTA}
\begin{align}
P_{\text{GUE}}(E_1,\dots,E_d)=c\prod_{i<j}^{1,\dots,d}(E_i-E_j)^2\exp\left(-\sum_{i=1}^dE_i^2\right),
\end{align}
\begin{figure}[t!]
\centering
\includegraphics[width=.48\textwidth]{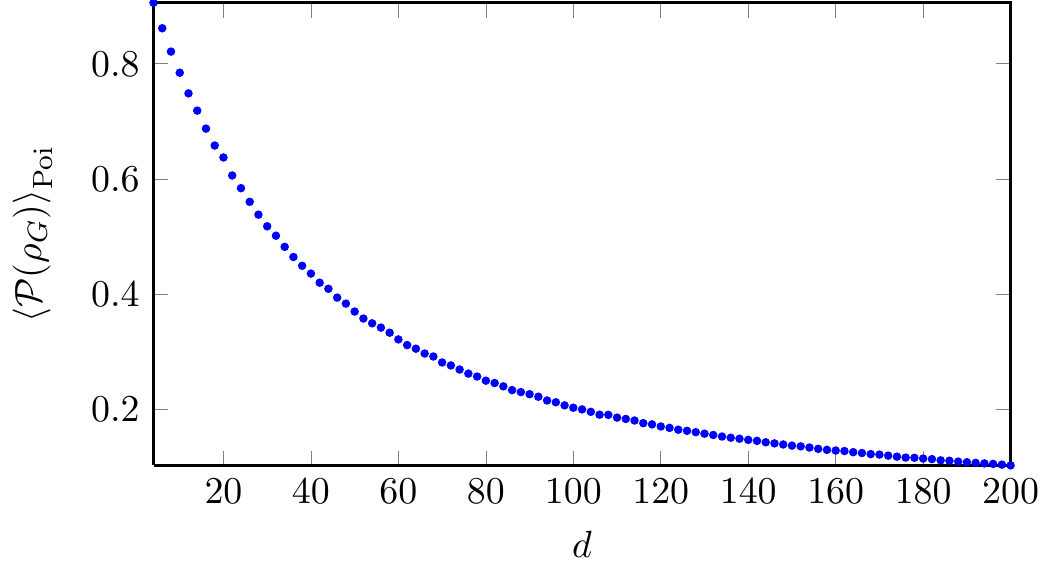}
\caption{(Color online) Average purity of the Gibbs state for regular systems with fixed value of $\beta=10$ as a function of $d$. The data was obtained by Monte Carlo sampling of energy values and each point corresponds to an average over $10,000$ samples.}
\label{fig.dedependence}
\end{figure}
with a normalization constant $c$. For both classes of systems this integral can be evaluated numerically for low dimensions. Furthermore, the absence of correlations in the regular spectrum allows to obtain average values of Eq.~(\ref{eq.averagegibbs}) even for higher dimensions by Monte Carlo sampling of equally distributed eigenenergies within the spectral span, which is determined by the normalization of the interaction strength. The dependence of $\left\langle\mathcal{P}(\rho_G)\right\rangle$ on $\beta$ is shown for $2\times2$-systems in Fig.~\ref{fig.betadependence} for both distributions and Fig.~\ref{fig.dedependence} shows how the average thermal purity of regular systems depends on $d_E$.

\subsection{Open system thermalization}
Next we want to refine our approach to the thermalization of quantum systems in contact with an environment \cite{LINDEN,HORODECKI}. To this end we deal with the term
\begin{align}
 &\left\|\text{Tr}_E\{\rho_G-\rho(t)\}\right\|^2\notag\\=\:&\left\|\text{Tr}_E\{W(e^{-\beta D}/Z)W^{\dagger}\}\right\|^2\notag\\&+\left\|\text{Tr}_E\{We^{-iDt}W^{\dagger}\rho(0)We^{iDt}W^{\dagger}\}\right\|^2\notag\\&-2\text{Tr}\left\{\text{Tr}_E\{W(e^{-\beta D}/Z)W^{\dagger}\}\right.\notag\\&\quad\times\left.\text{Tr}_E\{We^{-iDt}W^{\dagger}\rho(0)We^{iDt}W^{\dagger}\}\right\}.
\end{align}
After taking the unitary average of this quantity, we can identify the first two terms with already known expressions. Inserting $M=e^{-\beta D}/Z$ into Eq.~(\ref{eq.HSnormaverage}) yields
\begin{align}
 \left\langle\left\|\text{Tr}_E\{W(e^{-\beta D}/Z)W^{\dagger}\}\right\|^2\right\rangle=C_1\mathcal{P}(\rho_G)+C_2,
\end{align}
and from Eq.~(\ref{eq.generalresultcomplexsystem}) we obtain for $M=\rho(0)$,
\begin{align}
 &\left\langle\left\|\text{Tr}_E\{We^{-iDt}W^{\dagger}\rho(0)We^{iDt}W^{\dagger}\}\right\|^2\right\rangle\notag\\&=\tilde{c}_1(t)\mathcal{P}(\rho(0))+\tilde{c}_2(t)+\tilde{c}_3(t)\mathcal{P}(\rho_S(0))+\tilde{c}_4(t)\mathcal{P}(\rho_E(0)).
\end{align}
Averaging the third term requires knowledge of the sixth moment function of the unitary group:

\begin{widetext}
\begin{align}
&2\text{Tr}\left\{\text{Tr}_E\{W(e^{-\beta D}/Z)W^{\dagger}\}\text{Tr}_E\{We^{-iDt}W^{\dagger}\rho(0)We^{iDt}W^{\dagger}\}\right\}\notag\\
&=2\sum_{ijkl}\bra{\varphi_i\chi_j}W(e^{-\beta D}/Z)W^{\dagger}\ket{\varphi_k\chi_j}\bra{\varphi_k\chi_l}We^{-iDt}W^{\dagger}\rho(0)We^{iDt}W^{\dagger}\ket{\varphi_i\chi_l}\notag\\
&=2\sum_{ijl}\bra{\varphi_i\chi_j}\underbrace{W(e^{-\beta D}/Z)W^{\dagger}(I\otimes\ket{\chi_j}\bra{\chi_l})We^{-iDt}W^{\dagger}\rho(0)We^{iDt}W^{\dagger}}_{\langle\dots\rangle\rightarrow \mathbb{E}^{(6)}(e^{-\beta D}/Z,I\otimes\ket{\chi_j}\bra{\chi_l},e^{-iDt},\rho(0),e^{iDt})}\ket{\varphi_i\chi_l}.
\label{eq.sixthmomentapplied}
\end{align}
\end{widetext}

\begin{figure}[b!]
 \centering
  \includegraphics[width=.45\textwidth]{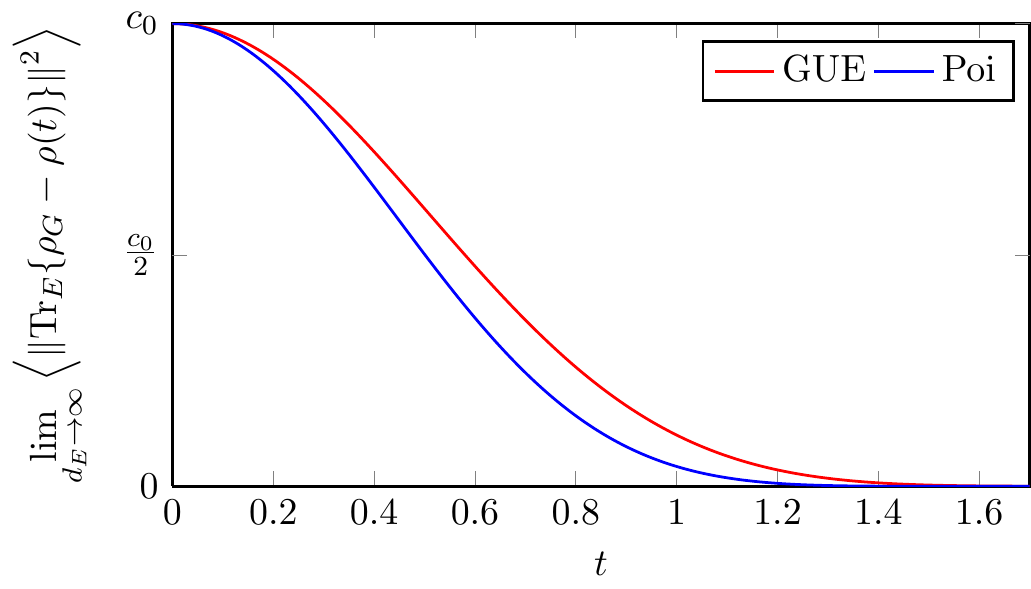}
 \caption{(Color online) Generic equilibration in the limit of infinite-dimensional environments for regular and chaotic systems.}
 \label{fig.EquiPoiGUElarged}
\end{figure}

The derivation of the general function $\mathbb{E}^{(6)}$ is given in the appendix \ref{ch.sixthmoment}. Here, we are particularly interested in the special case of $\mathbb{E}^{(6)}(e^{-\beta D}/Z,I\otimes\ket{\chi_j}\bra{\chi_l},e^{-iDt},\rho(0),e^{iDt})$, implying the choice of operators
\begin{align}
 &X_1=e^{-\beta D}/Z, X_2=I\otimes\ket{\chi_j}\bra{\chi_l},\notag\\&X_3=X_5^{\dagger}=e^{-iDt}, X_4=\rho(0),
\end{align}
which yields the simple result
\begin{align}
 &2\sum_{ijl}\bra{\varphi_i\chi_j}\notag\\&\times\mathbb{E}^{(6)}(e^{-\beta D}/Z,I\otimes\ket{\chi_j}\bra{\chi_l},e^{-iDt},\rho(0),e^{iDt})\ket{\varphi_i\chi_l}\notag\\&=2/d_S.\label{eq.crossterm}
\end{align}
Inserting this into Eq.~(\ref{eq.sixthmomentapplied}), we finally obtain
\begin{align}
 &\left\langle\left\|\text{Tr}_E\{\rho_G-\rho(t)\}\right\|^2\right\rangle=C_1\mathcal{P}(\rho_G)+C_2+\tilde{c}_1(t)\mathcal{P}(\rho(0))\notag\\&\qquad+\tilde{c}_2(t)+\tilde{c}_3(t)\mathcal{P}(\rho_S(0))+\tilde{c}_4(t)\mathcal{P}(\rho_E(0))-\frac{2}{d_S}.
\end{align}
The coefficients $C_1,C_2$ and $\tilde{c}_1(t),\dots,\tilde{c}_4(t)$ are well-known and the latter are also functions of $f(t)$. Averaging this result for certain eigenvalue distributions of the Hamiltonian again enables us to study the influence of level statistics and quantum chaos \cite{SREDNICKI}. We find remarkably simple expressions for the asymptotic values of large $d_E$:
\begin{align}
 \lim\limits_{d_E\rightarrow\infty}\left\langle\left\|\text{Tr}_E\{\rho_G-\rho(t)\}\right\|^2\right\rangle_{\text{Poi}}=c_0\left[\frac{\cos(t)\sin(t)}{t}\right]^4
\end{align}
and
\begin{align}
 \lim\limits_{d_E\rightarrow\infty}\left\langle\left\|\text{Tr}_E\{\rho_G-\rho(t)\}\right\|^2\right\rangle_{\text{GUE}}=c_0\left[\frac{J_1(2t)}{t}\right]^4,
\end{align}
with $c_0=\mathcal{P}(\rho_S(0))-1/d_S$. 
We see that both functions describe an oscillating, algebraic decay to zero. Thus,
for both eigenvalue distributions we find on average an asymptotic convergence to 
the thermal equilibrium state, see Fig.~\ref{fig.EquiPoiGUElarged}. While in the
case of Poissonian level statistics the average squared Hilbert-Schmidt distance
decays as $t^{-4}$, the same quantity behaves as $t^{-6}$ for the Gaussian 
unitary ensemble.

\section{Conclusions}
Employing general group theoretic results of Collins and \'{S}niady \cite{COLLINS} we have derived the explicit form of a class of operator-valued averages over the Haar measure. These were used to provide a tool-box for a variety of physically motivated applications. The thereby obtained expressions can be interpreted as averages over all possible time evolutions. Motivated by the foundations of statistical mechanics, we assume that the dynamics of generic complex quantum systems is well represented by the average value. This is endorsed explicitly for high-dimensional systems by our results, showing a vanishing variance in the limit of $d_E\rightarrow\infty$ which is also in agreement with concentration of measure arguments. Due to a generalization of the uniform result for general eigenvalue distributions of the total system, detailed analysis of the effect of level statistics can be performed using methods of random matrix theory.

To exemplify the general applicability and flexibility of our results we have investigated a series of topics associated with open quantum systems and related fields, which are of current interest. We obtain particularly simple results for the Hilbert-Schmidt distance of two quantum states in the open system, which was applied in the context of system-environment correlations. Our results suggest an observable signature of the total state correlations on a generic open-system evolution, which vanishes in the limit of large $d_E$.

Furthermore, we have investigated the generic evolution of system-environment entanglement in a pure total state. Using exact analytical expressions, we have found a universal evolution towards the maximally entangled state with vanishing variance in the limit of large environmental dimensions. These results were confirmed by an analysis of the generic time evolution for regular and chaotic systems.

Finally, we have studied the thermalization process of complex quantum systems. 
The average distance of the time-dependent reduced density operator to the 
reduced thermal equilibrium state was given as a function of time depending on 
system parameters and initial conditions. While generic closed systems never 
completely evolve from a non-equilibrium condition into equilibrium, 
high-dimensional environments lead to thermalization of generic open systems: 
As we have shown the average squared Hilbert-Schmidt distance
exhibits an algebraically decaying oscillation, where however regular and chaotic 
systems decay with different powers of time.

Using the results presented in this paper, expectation values for dynamical expressions of complex open systems can be investigated analytically. This may be especially useful if instead of the exemplary analysis of a particular model system, the expected performance of a method in generic systems is of interest.

\acknowledgments
M.G. thanks the German National Academic Foundation for support.

\begin{appendix}
\section{Moments of the unitary group}
In this section we summarize our findings on the unitary averages based on the general results of Collins and \'{S}niady \cite{COLLINS}. For additional details and terminology we refer the reader to the standard literature on group theory, see, e.g., Refs.~\cite{WIGNER,FULTON} and to the original publications on Weingarten functions \cite{COLLINSa,COLLINS,MATSUMOTO}. Reference \cite{COLLINS} provides a method to obtain exact expressions for average values of the form
\begin{align}
 \mathbb{E}^{(q+1)}(X_1,\dots,X_q)=\int d\mu(U) UX_1U^{\dagger}X_2U\dots UX_qU^{\dagger},
\end{align}
where $X_1,\dots,X_q$ are arbitrary operators on a $d$-dimensional Hilbert space $\mathcal{H}$ and $d\mu$ denotes the Haar measure on the unitary group $\mathcal{U}(d)$. The expression involves an integration over $n=q+1$ unitary matrices and is therefore called an $n$th moment function of the unitary group. While for higher-order terms we have to make use of the above result, one can obtain moments up to the order of $q=4$ also by elementary arguments using only the invariance property of the Haar measure, $d\mu(U)=d\mu(UV)=d\mu(VU),\forall\:U,V\in\mathcal{U}(d)$. The second moment function $\mathbb{E}^{(2)}(X)=\int d\mu(U) UXU^{\dagger}$ is a trivial consequence of this invariance property. In fact, $\mathbb{E}^{(2)}(X)$ commutes with every unitary matrix $V$ since 
\begin{eqnarray*}
 V\mathbb{E}^{(2)}(X)V^{\dagger} &=& 
 \int d\mu(VU) VUX(VU)^{\dagger} \\
 &=& \int d\mu(W) WXW^{\dagger} \\
 &=& \mathbb{E}^{(2)}(X).
\end{eqnarray*}
Using the irreducibility of the group of unitary matrices and Schur's Lemma we conclude that $\mathbb{E}^{(2)}(X)$ is proportional to the $d$-dimensional identity matrix: $\mathbb{E}^{(2)}(X)=c I$. The constant factor $c$ can be obtained by taking the trace of both expressions: $\text{Tr}\,\mathbb{E}^{(2)}(X)=cd=\text{Tr} X$, and we obtain the general form of the second moment function as $\mathbb{E}^{(2)}(X)=\text{Tr} X\cdot I/d$. An elementary derivation of the fourth moment function has been provided in Ref.~\cite{GB}.

\subsection{\texorpdfstring{The general expression for the $n$th moment}{The general expression for the nth moment}}

\begin{table*}[tb]
\centering
\hrulefill\\
\begin{minipage}{.45\textwidth}\centering
 \begin{tabular}{l|ccc}
  &(1,1,1)&(2,1)&(3)\\\hline
$\chi_{(3)}$ & 1 & 1 & 1\\
$\chi_{(2,1)}$ & 2 & 0 & -1\\
$\chi_{(1,1,1)}$ & 1 & -1 & 1
 \end{tabular}
\caption{Character table for the irreducible representations of $\mathcal{S}_3$ labeled by partitions of 3.}
\label{tab.charactersS3}
\end{minipage}\qquad
\begin{minipage}{.45\textwidth}\centering
  \begin{tabular}{l|ccc}
 $\lambda$ &(1,1,1)&(2,1)&(3)\\
$Z_{\lambda}$ & 1 & 3 & 2
 \end{tabular}
\caption{Number of constituents of the conjugacy classes of $\mathcal{S}_3$.}
\label{tab.elementsS3}
\end{minipage}\\
\hrulefill
\end{table*}

The elements of the matrix denoted by a capital letter will be denoted by the corresponding lowercase letter, leading to the correspondence of matrix elements $x^{\alpha}_{ij}$, $u_{ij}$ and matrices $X_{\alpha}$, $U$ respectively. We rewrite the expression for the $n$th moment function $\mathbb{E}^{(q+1)}(X_1,\dots,X_q)$ with $n=q+1$ in terms of the elements:
\begin{align}
 &\left(\mathbb{E}^{(q+1)}(X_1,\dots,X_q)\right)_{ij}\notag\\&=\sum_{k_1,\dots,k_{2q}}\left\langle u_{ik_1}x^1_{k_1k_2}u^{*}_{k_3k_2}\dots u_{k_{2q-2}k_{2q-1}}x^q_{k_{2q-1}k_{2q}}u^{*}_{jk_{2q}}\right\rangle\notag\\
&=\sum_{k_1,\dots,k_{2q}}x^1_{k_1k_2}\dots x^q_{k_{2q-1}k_{2q}}\notag\\&\qquad\times\left\langle u_{ik_1}\dots u_{k_{2q-2}k_{2q-1}} u^{*}_{k_3k_2}\dots u^{*}_{jk_{2q}}\right\rangle.\label{eq.eighthmomentfunction}
\end{align}
The expression $\langle u_{ik_1}\dots u_{k_{2q-2}k_{2q-1}} u^{*}_{k_3k_2}\dots u^{*}_{jk_{2q}}\rangle$ is called the $n$th moment of the unitary group, involving the average of a product of $n$ elements of a unitary matrix. Since all odd moments are equal to zero \cite{COLLINS}, we write $n=2m$. The result for the $n$th moment is a consequence of the Schur-Weyl duality \cite{COLLINS}:
\begin{align}
 &\left\langle u_{i_1j_1}\dots u_{i_mj_m}u^*_{i'_1j'_1}\dots u^*_{i'_mj'_m}\right\rangle\notag\\&=\sum_{\sigma,\tau\in\mathcal{S}_m}\delta_{i_1i'_{\sigma(1)}}\dots\delta_{i_mi'_{\sigma(m)}}\delta_{j_1j'_{\tau(1)}}\dots\delta_{j_mj'_{\tau(m)}}\text{Wg}\left(\tau\sigma^{-1}\right),\label{eq.collinstheorem}
\end{align}
where 
\begin{align}
 \text{Wg}(\sigma)=\frac{1}{(m!)^2}\sum_{\lambda\vdash m}\frac{\chi_{\lambda}(e)^2\chi_{\lambda}(\sigma)}{s_{\lambda,d}(1)}\label{eq.WGfunc}
\end{align}
denotes the Weingarten function, and the Schur function evaluated on $d$ copies of $x$ is written as $s_{\lambda,d}(x)=s_{\lambda,d}(x,\dots,x)$. The sum in Eq.~(\ref{eq.WGfunc}) is carried out over all partitions $\lambda$ of $m$, denoted by $\lambda\vdash m$. The characters of the irreducible representations of $\mathcal{S}_m$ labeled by partitions $\lambda$ and evaluated on the element $\sigma$ are represented by $\chi_{\lambda}(\sigma)$ and $e$ denotes the identity. The value of the Schur function for $x=1$ can be determined with the help of the following lemma \cite{COLLINS}: For any $\lambda\vdash m$ one has
\begin{align}
 s_{\lambda,d}(1)=\frac{1}{m!}\sum_{\tau\vdash m}d^{c(\tau)}\chi_{\lambda}(\tau)Z_{\tau},\label{eq.schurfunc}
\end{align}
where $c(\tau)$ denotes the number of cycles of $\tau$ and $Z_{\tau}$ the number of elements for the conjugacy class associated with $\tau$. With this result it is possible to obtain the exact expression for the $n$th moment of the unitary group given that the character table of $\mathcal{S}_m$ with respect to the irreducible characters labeled by partitions of $m$ and the number of constituents of the corresponding conjugacy classes are known.

\subsection{The fourth moment}
The fourth moment contains an integral over four unitary matrices, hence, according to Eq.~(\ref{eq.collinstheorem}), we have $m=2$. The conjugacy classes and characters of the symmetric group $\mathcal{S}_2$ are easily obtained since the group $\mathcal{S}_2$ only consists of two elements: The identity operation, corresponding to the partition $(1,1)$, and the swap operation with the partition $(2)$, each one forming its own conjugacy class. The two associated representations are the trivial representation and the sign representation. With Eqs.~(\ref{eq.WGfunc}) and (\ref{eq.schurfunc}), the two Weingarten functions are
\begin{align}
 \text{Wg}((1,1))&=\frac{1}{d^2-1},\notag\\
 \text{Wg}((2))&=-\frac{1}{d(d^2-1)}. 
\end{align}
With this we obtain the following average value with the aid of (\ref{eq.collinstheorem}),
\begin{align}
 \left(\mathbb{E}^{(4)}(X_1,X_2,X_3)\right)_{ij}=\sum_{k,\dots,p}x^1_{kl}x^2_{mn}x^3_{op}\left\langle u_{ik}u_{no}u^*_{ml}u^*_{jp}\right\rangle,
\end{align}
which yields:
\begin{align}
 \mathbb{E}^{(4)}(X_1,X_2,X_3)=\:&\frac{d\,\text{Tr} X_3X_1-\text{Tr} X_1\,\text{Tr} X_3}{d(d^2-1)}\,(\text{Tr} X_2) I\notag\\&+\frac{d\,\text{Tr} X_1\,\text{Tr} X_3-\text{Tr} X_3X_1}{d(d^2-1)}X_2.\label{eq.fourthmoment}
\end{align}
An alternative derivation can be found in Ref.~\cite{GB}.

\subsection{The sixth moment}
\label{ch.sixthmoment}
We derive the sixth moment function $(m=3)$ with the aid of the character table of $\mathcal{S}_3$ with respect to the irreducible characters labeled by partitions of 3, see Tab.~\ref{tab.charactersS3}, and the number of elements of all the three conjugacy classes, see Tab.~\ref{tab.elementsS3}. The Weingarten functions for the three conjugacy classes of $\mathcal{S}_3$ are
\begin{align}
\text{Wg}((1,1,1))=\frac{-2 + d^2}{d (4 - 5 d^2 + d^4)},\notag\\
\text{Wg}((2,1))=\frac{1}{-4 + 5 d^2 - d^4},\notag\\
\text{Wg}((3))=\frac{2}{4 d - 5 d^3 + d^5}.
\end{align}

The general sixth moment function is given in matrix elements as
\begin{align}
 &(\mathbb{E}^{(6)}(X_1,X_2,X_3,X_4,X_5))_{ij}\notag\\&=(\left\langle UX_1U^{\dagger}X_2UX_3U^{\dagger}X_4UX_5U^{\dagger} \right\rangle)_{ij}\notag\\
&=\left\langle\sum_{k,\dots,t}u_{ik}x^1_{kl}u^*_{ml}x^2_{mn}u_{no}x^3_{op}u^*_{qp}x^4_{qr}u_{rs}x^5_{st}u^*_{jt}\right\rangle\notag\\
&=\sum_{k,\dots,t}x^1_{kl}x^2_{mn}x^3_{op}x^4_{qr}x^5_{st}\left\langle u_{ik}u_{no}u_{rs}u^*_{ml}u^*_{qp}u^*_{jt}\right\rangle
\end{align}
The final result can be obtained with the aid of Eq.~(\ref{eq.collinstheorem}) and contains 36 terms, which we do not state explicitly.

\begin{table*}[tb]
\centering
\hrulefill\\
\begin{minipage}{.45\textwidth}\centering
 \begin{tabular}{l|ccccc}
 & (1,1,1,1) & (2,1,1) & (2,2) & (3,1) & (4)\\\hline
$\chi_{(4)}$ & 1 & 1 & 1 & 1 & 1\\
$\chi_{(3,1)}$ & 3 & 1 & -1 & 0 & -1\\
$\chi_{(2,2)}$ & 2 & 0 & 2 & -1 & 0\\
$\chi_{(2,1,1)}$ & 3 & -1 & -1 & 0 & 1\\
$\chi_{(1,1,1,1)}$ & 1 & -1 & 1 & 1 & -1\\
 \end{tabular}
\caption{Character table for the irreducible representations of $\mathcal{S}_4$ labeled by partitions of 4.}
\label{tab.characters4}
\end{minipage}\qquad
\begin{minipage}{.45\textwidth}\centering
\begin{tabular}{l|ccccc}
$\lambda$ & (1,1,1,1) & (2,1,1) & (2,2) & (3,1) & (4)\\
$Z_{\lambda}$ & 1 & 6 & 3 & 8 & 6\\
 \end{tabular}
\caption{Number of constituents of the conjugacy classes of $\mathcal{S}_4$.}
\label{tab.conjugacyclasses4}
\end{minipage}
\\\hrulefill
\end{table*}

\subsection{The eighth moment}
The eighth moment function ($m=4$), expressed element-wise reads
\begin{align}
 &\left(\mathbb{E}^{(8)}(X_1,\dots,X_7)\right)_{ij}\notag\\&=\sum\limits_{\substack{k,\dots,x\\\sigma,\tau\in\mathcal{S}_4}}x^1_{kl}x^2_{mn}x^3_{op}x^4_{qr}x^5_{st}x^6_{uv}x^7_{wx}\delta_{(\sigma,\tau)}\textrm{Wg}\left(\tau\sigma^{-1}\right)\label{eq.eighthmoment},
\end{align}
where
\begin{align}
\delta_{(\sigma,\tau)}=\delta_{i{\sigma(m)}}\delta_{n{\sigma(q)}}\delta_{r{\sigma(u)}}\delta_{v{\sigma(j)}}\delta_{k{\tau(l)}}\delta_{o{\tau(p)}}\delta_{s{\tau(t)}}\delta_{w{\tau(x)}}.
\end{align}
The Weingarten functions for the five conjugacy classes of $\mathcal{S}_4$ can be obtained using Tabs.~\ref{tab.characters4} and \ref{tab.conjugacyclasses4}:
\begin{align}
 \text{Wg}((4)) &= \frac{-5}{ad},\notag\\
 \text{Wg}((3,1)) &= \frac{-3 + 2 d^2}{ad^2},\notag\\
\text{Wg}((2,2)) &= \frac{6 + d^2}{ad^2},\notag\\
\text{Wg}((2,1,1)) &= \frac{-1}{9 d - 10 d^3 + d^5},\notag\\
\text{Wg}((1,1,1,1)) &= \frac{6 - 8 d^2 + d^4}{ad^2},
\end{align}
with $a=-36 + 49 d^2 - 14 d^4 + d^6$. The sum of Eq.~(\ref{eq.eighthmoment}), generally containing $4!^2=576$ terms, reduces to a simpler expression if the operators $X_1,\dots,X_7$ are explicitly chosen for physical applications.

\section{\texorpdfstring{Averages of functions of $f(t)$}{Averages of functions of f(t)}}
\label{ch.rmt}
\subsection{Regular systems}
\label{ch.regularsystemsaverage}
Regular systems are characterized by Poissonian level spacings statistics \cite{HAAKE,MEHTA}. The energy levels are randomly distributed and show no correlations. In this appendix we derive the average values of functions of the Fourier transform of the level density which continuously appear throughout the manuscript. We start with
\begin{align}
 \langle|f(t)|^2\rangle=&\frac{1}{d^2}\left\langle\sum_{i,j}e^{-i (E_i-E_j)t}\right\rangle\notag\\
=&\frac{1}{d}+\frac{1}{d^2}\int dE_1\int dE_2e^{-i (E_1-E_2)t}\notag\\&\times\underbrace{\sum_{\substack{i,j\\(i\neq j)}}\left\langle\delta(E_1-E_i)\delta(E_2-E_j)\right\rangle}_{R_2(E_1,E_2)}.\label{eq.meanfsquare}
\end{align}
Since there are no correlations in the energy levels for Poissonian statistics, the 2-point correlation function is given by $R_2(E_1,E_2)=R_1(E_1)R_1(E_2)$. The level density $R_1(E)$ is flat and homogeneous. Fixing the size of the matrix elements of the Hamiltonian by means of
$\langle|H_{ij}|^2\rangle=1/d$ yields $R_1(E)=\frac{d}{4}\Theta(E+2)\Theta(2-E)$ \cite{GARCIAMATA}. The resulting average of $|f(t)|^2$ is given by
\begin{align}
 \langle|f(t)|^2\rangle_{\text{Poi}}=\frac{1}{d}+\frac{d-1}{d}\left[\frac{\sin(2t)}{2t}\right]^2,
\end{align}
and further average values, in particular those of the functions $\Re\{f(t)^2f^*(2t)\}$ and $|f(t)|^4$ can be derived by analogous methods. We obtain
\begin{widetext}
 \begin{align}
\langle\Re\{f(t)^2f^*(2t)\}\rangle_{\text{Poi}}=\frac{1}{d^2} + \frac{d-1}{d^2}\left[\frac{\sin(4t)}{4t}\right]^2 + 2\frac{d-1}{d^2}\left[\frac{\sin(2t)}{2t}\right]^2+\frac{(d-2)(d-1)}{d^2}\cos(2t)\left[\frac{\sin(2t)}{2t}\right]^3
\end{align}
and
\begin{align}
\langle|f(t)|^4\rangle_{\text{Poi}}=\frac{3d-2}{d^3}+\frac{4(d-1)^2}{d^3}\left[\frac{\sin(2t)}{2t}\right]^2+\frac{2(d-1)(d-2)}{d^3}\cos(2t)\left[\frac{\sin(2t)}{2t}\right]^3+\frac{(d-1)(d-2)(d-3)}{d^3}\left[\frac{\sin(2t)}{2t}\right]^4.
\end{align}
\end{widetext}

\subsection{Chaotic systems: Gaussian unitary ensemble}
\label{ch.gue}
In this section we average the same functions of $f(t)$ but now with respect to the GUE, representing chaotic systems with no time-reversal symmetry. We can build on the general results of the previous section using the $n$-point correlation functions for the GUE \cite{MEHTA},
\begin{align}
 R_n(E_1,\dots,E_n)=\text{det}\left[\sum_{k=0}^{d-1}\varphi_k(E_i)\varphi_k(E_j)\right]_{i,j=1,\dots,n},
\end{align}
where
\begin{align}
 \varphi_k(x)=(2^kk!\sqrt{2\pi/d})^{-1/2}\exp\left(-\frac{d}{4}x^2\right)\mathsf{H}_k\left(\sqrt{\frac{d}{2}}x\right),
\end{align}
and
\begin{align}
 \mathsf{H}_k(x)=\exp(x^2)\left(-\frac{d}{dx}\right)^k\exp(-x^2)
\end{align}
denotes the Hermite polynomial of order $k$. The average value of $|f(t)|^2$ is then obtained from Eq.~(\ref{eq.meanfsquare}). The average values of $\Re\{f(t)^2f^*(2t)\}$ and $|f(t)|^4$ are determined analogously. The integrations for the latter can be carried out with standard mathematical software but become fairly intricate for larger values of $d$. In fact, so as to facilitate computational analysis, it is possible to replace $\langle|f(t)|^4\rangle\approx\langle|f(t)|^2\rangle^2$. Correlations of fourth order are approximated by second order terms and the expression can be considered as almost exact for our purposes. A more drastic approximation which allows analytical expressions for large values of $d$ can be performed by neglecting all correlations and interchanging averages and powers of the function $f(t)$ \cite{GARCIAMATA}. For the GUE this is valid in the leading order of $1/d$ \cite{GARCIAMATA}. Introducing $h(t)\equiv\langle f(t)\rangle$ the three functions of interest are approximated by \cite{GARCIAMATA}
\begin{align}
 \langle|f(t)|^2\rangle&\approx|h(t)|^2,\notag\\
 \langle\Re\{f(t)^2f^*(2t)\}\rangle&\approx\Re\{h(t)^2h^*(2t)\},\notag\\
 \langle|f(t)|^4\rangle&\approx|h(t)|^4.
 \end{align}
The function $h(t)$ is given by
\begin{align}
 h(t)&=\frac{1}{d}\left\langle\sum_{j=1}^{d}e^{-i E_jt}\right\rangle=\frac{1}{d}\int dEe^{-i Et}\underbrace{\sum_{j=1}^{d}\left\langle\delta(E-E_j)\right\rangle}_{R_1(E)}.
\label{eq.hoft}
\end{align}
For the GUE this yields \cite{GARCIAMATA}
\begin{align}
 h_{\text{GUE}}(t)=\frac{1}{d}\int dEe^{-i Et}\sum_{k=0}^{d-1}\varphi^2_k(E)\xrightarrow{\text{large }d}J_1(2t)/t,
\end{align}
where $J_1(t)$ denotes the Bessel function of the first kind.
\end{appendix}

\end{document}